\documentclass[lettersize,journal]{IEEEtran}
\IEEEoverridecommandlockouts
\usepackage{cite}
\usepackage[cmex10]{amsmath}
\usepackage{amsthm}
\usepackage{amssymb}
\usepackage{balance}
\usepackage[final]{graphicx}
\usepackage{color}
\usepackage{epsfig}
\usepackage{epstopdf}
\usepackage{tikz}
\usepackage{comment}
\usepackage{subfiles}
\usetikzlibrary{positioning}
\usepackage[ruled,vlined,linesnumbered,noend]{algorithm2e}

\newtheorem{lemma}{Lemma}

\usepackage{bbm}

\newcommand{\bs}[1]{\boldsymbol{#1}}
\newcommand{\mc}[1]{\mathcal{#1}}

\newcommand{\mb}[1]{\mathbf{#1}}
\newcommand{\mr}[1]{\mathrm{#1}}
\newcommand{\ms}[1]{\mathsf{#1}}
\newcommand{\tr}{\mathrm{Tr}}

\newcommand{\lr}[1]{\langle #1 \rangle}
\newcommand{\blr}[1]{\big\langle #1 \big\rangle}
\newcommand{\bblr}[1]{\bigg\langle #1 \bigg\rangle}

\DeclareMathOperator*{\argmax}{arg\;max}

\definecolor{LayerColor}{RGB}{230, 230, 230}
\definecolor{InOutColor}{RGB}{240, 243, 255}
\definecolor{cellColor}{RGB}{230, 230, 230}

\begin{document}
	
	\title{Variational Bayesian Channel Estimation and Data Detection for Cell-Free Massive MIMO with Low-Resolution Quantized Fronthaul Links 
	
	\author{Sajjad Nassirpour, \emph{Member, IEEE}, Toan-Van Nguyen, \emph{Member, IEEE}, Hien Quoc Ngo, \emph{Fellow, IEEE}, Le-Nam Tran, \emph{Senior Member, IEEE}, Tharmalingam Ratnarajah, \emph{Senior Member, IEEE}, and Duy H. N. Nguyen, \emph{Senior Member, IEEE}
	\vspace{-5mm}
    
 % \author{\IEEEauthorblockN{Sajjad Nassirpour, Toan-Van Nguyen, and Duy H. N. Nguyen}

% \IEEEauthorblockA{\textit{Department of Electrical and Computer
% Engineering, San Diego State University, San Diego, USA}\\
% Email: snassirpour@sdsu.edu, , and }

\thanks{S. Nassirpour, T. V. Nguyen, T. Ratnarajah, and D. H. N. Nguyen are with the Department of Electrical and Computer
Engineering, San Diego State University, San Diego, CA 92182, USA. Emails: {\sffamily snassirpour@sdsu.edu}, {\sffamily tnguyen58@sdsu.edu}, {\sffamily tratnarajah@sdsu.edu}, and {\sffamily duy.nguyen@sdsu.edu}.

H. Q. Ngo is with the School of Electronics, Electrical Engineering and Computer Science, Queen's University,  Belfast, UK. Email:
{\sffamily hien.ngo@qub.ac.uk}. 

L. N. Tran is with the
School of Electrical and Electronic Engineering, University College Dublin, Dublin, Ireland. Email:
{\sffamily nam.tran@ucd.ie}.}

%\thanks{This work was supported in part by the U.S. National Science Foundation under grants ECCS-2146436, CCF-2225576, and the U.S.-Ireland R\&D Partnership Program under grant CCF-2322190; the U.K. Research and Innovation Future Leaders Fellowships under Grant MR/X010635/1; and a research grant from the Department for the Economy Northern Ireland under the U.S.-Ireland R\&D Partnership Programme. The work of L.-N. Tran was supported in part by Taighde \'{E}ireann - Research Ireland under Grant numbers 22/US/3847 and 13/RC/2077 P2.}

}
}
	
	\maketitle
	\begin{abstract} 
        % We study the joint channel estimation and data detection (JED) problem in a cell-free massive multiple-input multiple-output (CF-mMIMO) network, where access points (APs) communicate with a central processing unit (CPU) over fronthaul links. However, the bandwidth of these links is limited, and thus, presents challenges to the applicability of CF-mMIMO, especially with an ever-increasing number of users. To address this, we propose a method based on variational Bayesian (VB) inference for performing the JED process, where the APs forward low-resolution quantized versions of the signals to the CPU. We consider two approaches: \emph{quantization-and-estimation} (Q-E) and \emph{estimation-and-quantization} (E-Q). {\color{blue}In the Q-E approach, each AP uses low-bit quantizers to quantize its received signals before forwarding them to the CPU. In contrast, in the E-Q approach, each AP first performs local channel estimation during the pilot transmission phase and then sends a quantized version of its locally estimated channel, along with a quantized version of the received signals during the data transmission phase, to the CPU. The final JED process in both Q-E and E-Q is performed at the CPU.}
We study the joint channel estimation and data detection (JED) problem in cell-free massive MIMO (CF-mMIMO) networks, where access points (APs) forward signals to a central processing unit (CPU) over fronthaul links. Due to bandwidth limitations of these links, especially with a growing number of users, efficient processing becomes challenging. To address this, we propose a variational Bayesian (VB) inference-based method for JED that accommodates low-resolution quantized signals from APs. We consider two approaches: \emph{quantization-and-estimation} (Q-E) and \emph{estimation-and-quantization} (E-Q). In Q-E, each AP directly quantizes its received signals before forwarding them to the CPU. In E-Q, each AP first estimates channels locally during the pilot phase, then sends quantized versions of both the local channel estimates and received data to the CPU. The final JED process in both Q-E and E-Q is performed at the CPU.
% We evaluate the performance of our VB-based approach under three scenarios: perfect fronthaul link (PFL) with unquantized received signals, Q-E, and E-Q. The evaluation is conducted in terms of symbol error rate (SER), normalized mean square error (NMSE) of channel estimation, computational complexity, and fronthaul signaling overhead. {\color{blue}We compare our VB-based methods against both linear and nonlinear state-of-the-art JED approaches. Numerical results show that all of our VB-based methods outperform the linear JED baseline by leveraging the nonlinear modeling capabilities of VB. Moreover, our methods also surpass existing nonlinear counterparts, owing to a fully VB-driven framework where all processing is handled within the VB formulation. This allows convergence to a local optimum even under correlated channels, whereas other methods may diverge or rely on hybrid techniques, such as combining VB with expectation maximization, that can degrade performance.}%Furthermore, the VB(Q-E) method outperforms VB(E-Q) due to errors in the local channel estimation process at the APs within the VB(E-Q) approach.
We evaluate our proposed approach under perfect fronthaul links (PFL) with unquantized received signals, Q-E, and E-Q, using symbol error rate (SER), channel normalized mean squared error (NMSE), computational complexity, and fronthaul signaling overhead as performance metrics. Our methods are benchmarked against both linear and nonlinear state-of-the-art JED techniques. Numerical results demonstrate that our VB-based approaches consistently outperform the linear baseline by leveraging the nonlinear VB framework. They also surpass existing nonlinear methods due to:
\emph{i)} a fully VB-driven formulation, which performs better than hybrid schemes such as VB combined with expectation maximization; and
\emph{ii)} the stability of our approach under correlated channels, where competing methods may fail to converge or experience performance degradation.
	\end{abstract}
	%%%%%%%%%%%%%%%%%%%%%%%%%%%%%%%%%%%%%%%%%%%%%%%%%%%
	
	%%%%%%%%%%%%%%%%%%%%%%%%%%%%%%%%%%%%%%%%%%%%%%%%%%%
	\section{Introduction}
        \label{Sec:intro}
        Massive multiple-input multiple-output (mMIMO) is a key technology for improving spectral efficiency and interference management in fifth-generation (5G) and beyond wireless networks. By exploiting spatial diversity, mMIMO supports multiple users within the same frequency-time block \cite{boccardi2014five,tan2017spectral,lu2014overview, nassirpour2022power,gupta2023greenmo}. However, its performance may degrade under certain conditions: \emph{i.)} in highly correlated channels, where rank deficiency limits coverage; \emph{ii.)} with widely distributed users, where large user-base station (BS) distances cause significant path loss; and \emph{iii.)} at cell edges, where users experience strong inter-cell interference from neighboring mMIMO BSs.

% Massive multiple-input multiple-output (mMIMO) is a cornerstone technology for enhancing spectral efficiency and interference management in fifth-generation (5G) and beyond wireless networks. By exploiting spatial diversity, mMIMO enables multiple users to be supported within the same frequency-time block \cite{boccardi2014five,tan2017spectral,lu2014overview, nassirpour2022power,gupta2023greenmo}. However, its performance may deteriorate under specific conditions: \emph{i.)} in networks with highly correlated channels, where rank deficiency in the channel matrix can limit coverage; \emph{ii.)} in scenarios with widely distributed users, where the large distances between users and the mMIMO base station (BS) result in significant path loss, thereby reducing spectral efficiency; and \emph{iii.)} in cellular networks with cell-edge users, who are subject to considerable interference from multiple mMIMO BSs operating in neighboring cells.

To tackle the above challenges, cell-free mMIMO (CF-mMIMO) has attracted growing interest from both academia and industry \cite{buzzi2017cell,bjornson2020scalable,ngo2017total,ngo2017cell, nayebi2017precoding,bjornson2019making}. Unlike traditional mMIMO, which uses a large number of collocated antenna elements at a single BS, CF-mMIMO employs many distributed access points (APs). Each AP may be equipped with a single or multiple antenna elements, and they are positioned far apart from one another. This distributed architecture reduces channel correlation and lowers average path loss between users and APs compared to the collocated antenna setup in traditional mMIMO.

% To tackle the aforementioned challenges, the concept of cell-free mMIMO (CF-mMIMO) has garnered significant attention from both academia and industry \cite{buzzi2017cell,bjornson2020scalable,ngo2017total,ngo2017cell, nayebi2017precoding,bjornson2019making}. Unlike traditional mMIMO networks, where a large number of collocated antenna elements are deployed at a single BS, CF-mMIMO consists of many distributed access points (APs). Each AP may be equipped with a single or multiple antenna elements, and they are positioned far apart from one another. This distributed design reduces channel correlation compared to traditional mMIMO networks. Additionally, the separation of antenna elements results in lower average path loss between users and APs compared to the collocated antenna setup in traditional mMIMO networks.

Moreover, CF-mMIMO adopts a user-centric design, unlike the cell-centric approach of traditional mMIMO where a single BS serves all users in a cell. In CF-mMIMO, distributed APs coordinate with a central processing unit (CPU) via fronthaul links, enhancing connectivity and overall system performance.

% Moreover, CF-mMIMO employs a user-centric design paradigm, in contrast to the cell-centric design of traditional mMIMO networks, where all users in a cell are managed by a single BS. In CF-mMIMO, the focus is on maintaining consistent wireless link quality for both downlink and uplink connections. Each AP communicates with a central processing unit (CPU) through fronthaul links for coordination, which enhances system performance, especially network connectivity. 

Initial CF-mMIMO studies assumed fully centralized processing at the CPU \cite{ngo2017cell, nayebi2017precoding}, showing notable gains in median and $95 \%$-likely spectral efficiency over traditional small-cell mMIMO, where each BS antenna serves its own users. However, this approach demands high fronthaul bandwidth, as each AP must employ a high-resolution quantizer (e.g., $10+$ bits) to generate nearly continuous signals for transmission to the CPU. This becomes increasingly unsustainable as the number of users in 5G and beyond networks continues to grow. To mitigate this, the authors of \cite{bjornson2019making} proposed local processing at APs, enabling partial to fully decentralized architectures. Their approach used a linear filter (i.e., the linear minimum mean squared error (LMMSE) filter) to enhance spectral efficiency under limited fronthaul bandwidth.

% Initial studies on CF-mMIMO assumed that all processing is centralized at a CPU \cite{ngo2017cell, nayebi2017precoding}. These studies demonstrated significant improvements in both median and $95 \%$-likely spectral efficiency compared to traditional mMIMO in small-cell networks, where each antenna element at the BS exclusively serves its own set of users. However, centralized processing requires substantial fronthaul bandwidth. Specifically, each AP must employ a high-resolution quantizer (e.g., $10+$ bits) to generate nearly continuous signals for transmission to the CPU, which becomes increasingly unsustainable as the number of users in 5G and beyond networks continues to grow. To address the limited bandwidth challenge, the authors of \cite{bjornson2019making} proposed incorporating local processing at the APs, enabling portions of the processing to be performed locally. This study introduced three levels of local processing, ranging from partial to fully decentralized processing, and achieved improved spectral efficiency using linear filters, particularly the linear minimum mean squared error (LMMSE) filter.

While linear filters offer low computational complexity, their performance declines when the number of AP antennas is small relative to the number of users or when channels are highly correlated. To address these limitations, nonlinear methods such as approximate message passing (AMP), generalized AMP (GAMP), expectation propagation (EP), and variational Bayesian (VB) inference have been proposed \cite{donoho2009message,jeon2015optimality,nguyen2022variational,xiong2017low,karataev2024bilinear}. AMP is effective for data detection in mMIMO under independent and identically distributed (i.i.d.) Rayleigh fading by decoupling the system into parallel additive white Gaussian noise (AWGN) channels \cite{donoho2009message,jeon2015optimality}. However, it may diverge under ill-conditioned or non-zero-mean channels (e.g., correlated or Rician fading) \cite{nguyen2022variational}. Moreover, GAMP extends AMP to support non-Gaussian priors and quantized measurements \cite{xiong2017low}; however, it suffers instability with non-i.i.d. measurement matrices or small system sizes. On the other hand, EP minimizes Kullback-Leibler (KL) divergence via message passing and moment matching \cite{karataev2024bilinear}, but it lacks convergence guarantees and may diverge in high-dimensional settings.

The performance of CF-mMIMO networks strongly depends on accurate channel estimation (CE), motivating several studies on CE techniques \cite{zhong2024subspace,jin2019channel}. For instance, \cite{zhong2024subspace} proposed a subspace-based semi-blind CE scheme to mitigate pilot contamination, which arises when the number of users exceeds the available pilot duration. To further address this issue, \cite{liu2020graph} developed a graph coloring-based pilot assignment method that models user interference through AP selection, improving pilot reuse and reducing contamination. Beyond traditional CE, recent works have explored joint channel estimation and data detection (JED), where data detected during the data transmission phase is used to refine CE performance. In \cite{garg2022generalized}, a generalized superimposed pilot scheme was introduced for JED, distributing data symbols across coherence time slots via linear precoders. Then, \cite{song2021joint} proposed an iterative JED algorithm to solve a relaxed maximum a posteriori (MAP) problem using the forward-backward splitting technique.

Motivated by the above, in this paper, we focus on the JED problem in the uplink of CF-mMIMO networks and propose an approach based on VB inference. The VB method offers a robust framework for approximating posterior distributions by optimizing a simpler, tractable distribution to closely match the intractable true posterior. Originally developed for machine learning, VB techniques have recently gained attraction in wireless communications \cite{thoota2021variational, LyVannguyen2022variational, nguyen2022variational, wan2022joint}. Unlike other nonlinear methods like AMP, GAMP, and EP, VB is well-suited for ill-conditioned channels (e.g., correlated channels) and has solid convergence properties \cite{Bishop-2006, Wainwright-2008}.

% Motivated by the above, in this paper, we focus on the JED problem in the uplink scenario of CF-mMIMO networks and propose an approach based on VB inference. The VB method provides a robust statistical inference framework for approximating the posterior distributions of latent variables. This is achieved by identifying and optimizing a simpler, tractable distribution within a predefined family to closely approximate the intractable true posterior. Initially developed within the machine learning community, VB techniques have recently found significant applications in wireless communications~\cite{thoota2021variational,LyVannguyen2022variational,nguyen2022variational,wan2022joint}. {\color{blue}The VB method, unlike other nonlinear statistical methods such as AMP, GAMP, and EP, is well-suited for handling ill-conditioned channels (e.g., correlated channels) and has solid convergence properties \cite{Bishop-2006, Wainwright-2008}.}

It is worth noting that recent studies in \cite{guo2021joint, wang2020grant, iimori2021grant} have explored joint user activity detection and channel estimation in CF-mMIMO networks, formulating the problem as a compressed sensing (CS) task by leveraging user sparsity. Specifically, \cite{guo2021joint} employed a single measurement vector (SMV)-based minimum mean squared error (MMSE) approach, while \cite{wang2020grant} used the GAMP algorithm to tackle the CS problem. In \cite{iimori2021grant}, the authors extended the problem to include data detection and proposed a grant-free scheme based on bilinear inference, using the bilinear Gaussian belief propagation (BiGaBP) algorithm. Their method achieved efficient joint estimation without data spreading and mitigated pilot contamination through a low-coherence pilot design.

% It is worth noting that recent studies in \cite{guo2021joint,  wang2020grant, iimori2021grant} have investigated joint user activity detection and channel estimation in CF-mMIMO networks. These works formulated the problem as a compressed sensing (CS) task, utilizing the sparse activity model of users. The studies in \cite{guo2021joint, wang2020grant} applied techniques such as a single measurement vector (SMV)-based minimum mean squared error (MMSE) approach in \cite{guo2021joint} and the GAMP algorithm in \cite{wang2020grant}. {\color{blue}Then, the authors in \cite{iimori2021grant} focused on the joint user activity, channel estimation, and data detection problem in CF-mMIMO systems and proposed a grant-free access scheme based on bilinear inference, utilizing an activity-aware bilinear Gaussian belief propagation (BiGaBP) algorithm. The approach achieved efficient joint estimation without relying on data spreading, and employed a low-coherence pilot design based on frame theory to mitigate pilot contamination and reduce overhead.}

To address the limited bandwidth of fronthaul links, in addition to the previously discussed approach of local processing at the APs, two alternative methods can be considered. The first, called the \emph{quantization-and-estimation} (Q-E) scenario, uses low-bit quantizers at each AP to send quantized received signals to the CPU. The second, referred to as the \emph{estimation-and-quantization} (E-Q) scenario, performs local CE at each AP during the pilot transmission phase and forwards quantized versions of the local CE and received signals during the data transmission phase to the CPU. In both cases, final JED is performed at the CPU. Unlike the perfect fronthaul link (PFL) scenario with unquantized received signals, the CPU in Q-E and E-Q relies on quantized inputs, introducing challenges due to the nonlinearity of the quantization function and the resulting quantization noise. In \cite{bashar2020uplink}, the authors studied spectral and energy efficiency under Q-E and E-Q, using uniform quantization modeled by the Max algorithm and Bussgang decomposition.

Next, the work in \cite{zhao2024joint} focused on joint user activity detection and channel estimation in a CF-mMIMO network under Q-E with mixed-quantization APs, where some APs receive unquantized signals while others use low-bit quantizers. This problem was modeled as a CS task, and the authors proposed an approach based on multiple measurement vectors and GAMP to tackle it. Then, the study in \cite{takahashi2022bayesian} addressed the JED problem in the CF-mMIMO network under Q-E. The authors proposed a two-stage solution: first, a de-quantization step based on scalar Gaussian approximation and Bussgang decomposition to estimate the statistics of the unquantized signals; second, they applied a BiGaBP algorithm for JED. Nevertheless, the Bussgang decomposition discussed in \cite{zhao2024joint, takahashi2022bayesian}, which linearizes the relationship between quantized and unquantized signals, has two key limitations: it requires access to signal statistics before and after quantization, and it is applicable only when the input to the quantizer follows a Gaussian distribution. These constraints limit its applicability and leave room for improvements beyond Bussgang-based methods \cite{jacobsson2017throughput}. Building on this idea, \cite{wen2015bayes} studied the JED problem in mMIMO systems with quantized signals. The authors used a variant of belief propagation (BP) within a GAMP-based framework to approximate the marginal distributions of data and channel components, and derived analytical expressions for quantized observations using the Gaussian cumulative distribution function (CDF). However, this method may diverge under correlated channels, a known limitation of AMP-type algorithms. To address this, \cite{nguyen2024variational} proposed a VB-based JED framework for mMIMO systems and used the expectation maximization (EM) technique to estimate the precision terms. The VB framework guarantees convergence to at least a local optimum \cite{Bishop-2006, Wainwright-2008}.

Motivated by the results in \cite{nguyen2024variational}, in this work, we propose a VB-based approach to model the nonlinear relationship between unquantized and quantized signals in both Q-E and E-Q scenarios. Furthermore, we adopt Gamma priors for the precision parameters, generalizing the VB framework and showing that the VB-EM approach is a special case. %Our use of Gamma priors is inspired by related techniques in CS \cite{babacan2009bayesian}, where such priors are commonly used for modeling the precision of Gaussian variables. 
We evaluate the performance of our proposed VB-based method in terms of symbol error rate (SER), normalized mean squared error (NMSE) for channel estimation, computational complexity, and fronthaul signaling overhead.

% Motivated by the results in \cite{nguyen2024variational}, in this paper, we propose a VB-based approach to approximate the relationship between the unquantized signals and their quantized counterparts using a nonlinear Bayesian statistical model in both the Q-E and E-Q scenarios. Furthermore, we model the precision parameters using Gamma distributions, which generalizes the framework and enables us to view the EM approach as a special case of our Gamma-prior-based method. This provides a more flexible and theoretically sound foundation. We compare the performance of our VB-based methods against these two scenarios in terms of symbol error rate (SER), normalized mean square error (NMSE) for channel estimation, computational complexity, and fronthaul signaling overhead.}

{\bf Contribution:}
Our main contributions are as follows:

\begin{itemize}
    \item We consider uplink communications in a CF-mMIMO network and propose a method based on VB inference to address the JED problem in both Q-E and E-Q scenarios.

\item Unlike previous works in \cite{zhao2024joint,takahashi2022bayesian} that utilized Bussgang decomposition to linearize the relationship between quantized and unquantized signals, we leverage the VB framework to approximate the posterior distribution of the unquantized signal given its quantized counterpart.

\item We model uncertainty in the JED process via residual inter-user interference, assumed to be Gaussian, capturing both noise and detection/estimation errors. Its precision is also treated as a random variable and estimated within the VB framework.
%We model the uncertainty in the JED process using residual inter-user interference, which captures both the impact of noise and errors during the JED process. We assume that the residual inter-user interference follows a Gaussian distribution. Additionally, we treat its precision as a random variable and allow the VB framework to estimate it.}% Our numerical results show that the proposed VB method outperforms its variant, introduced in \cite{nguyen2024variational}, where the precision is estimated using the EM technique, referred to as the VB-EM method.}

\item We assess the performance of our proposed VB-based approach under the PFL, Q-E, and E-Q scenarios in terms of SER, channel NMSE, computational complexity, and fronthaul signaling overhead. We compare its performance with LMMSE(PFL), GAMP(PFL), GAMP(Q-E) \cite{wen2015bayes}, VB-EM(PFL), and VB-EM(Q-E) \cite{nguyen2024variational}, where VB-EM is built upon the VB framework and estimates the precision of the residual inter-user interference using EM. Numerical results show that our VB-based methods consistently outperform LMMSE(PFL) by capturing nonlinear relationships, and also surpass the GAMP- and VB-EM-based methods, thanks to a unified, fully VB-driven design that ensures convergence even under correlated channels. In contrast, GAMP may diverge, and EM-based precision estimation within VB-EM can degrade performance. Finally, VB(Q-E) slightly outperforms VB(E-Q) due to local CE errors in the latter, aligning with the findings in \cite{bashar2020uplink}.
%\item We assess the performance of the proposed VB-based approach under the PFL, Q-E, and E-Q scenarios in terms of SER, channel NMSE, computational complexity, and fronthaul signaling overhead. Further, we compare the performance of our VB-based approaches with LMMSE(PFL), GAMP(PFL), GAMP(Q-E) \cite{wen2015bayes}, VB-EM(PFL), and VB-EM(Q-E) \cite{nguyen2024variational}. The VB-EM method is built upon the VB framework, where the precision of the residual inter-user interference is estimated using the EM technique. The results show that our proposed VB-based methods consistently outperform LMMSE(PFL) by effectively capturing nonlinear relationships through the VB framework. In addition, our VB method achieves superior performance compared to GAMP and VB-EM, thanks to a unified and purely VB-driven design in which all inference steps are conducted within the VB paradigm. This integrated approach enhances robustness and ensures convergence to a local optimum, even in correlated channels. In contrast, GAMP methods may suffer from divergence issues, and VB-EM relies on EM to estimate the precision terms, which can compromise performance. Finally, the VB(Q-E) approach slightly outperforms the VB(E-Q) method due to the errors in the local CE process within the VB(E-Q) method, which is consistent with the findings in \cite{bashar2020uplink}.
\end{itemize}

The rest of this paper is structured as follows. Section~\ref{Sec:system_model} introduces the system model. Section~\ref{Sec:VB_for_JED} details the proposed VB framework for the JED problem. Section~\ref{Sec:numerical} presents the simulation results, and Section~\ref{Sec:conclusion} provides the conclusion.

{\bf Notation:}
In this paper, scalars are represented by italic letters, vectors by bold lowercase letters, and matrices by bold uppercase letters. The notation $\mc{CN}(\mb{m}, \mb{C})$ denotes a complex Gaussian random vector with mean $\mb{m}$ and covariance $\mb{C}$. $\Gamma(a, b)$ implies a Gamma distribution parametrized by $a$ and $b$. The space of $x \times y$ complex-valued matrices is denoted as $\mathbb{C}^{x \times y}$. We use $\mr{diag}\{\mb{v}\}$ to represent a diagonal matrix formed from the vector $\mathbf{v}$. Further, we indicate the determinant, transpose, conjugate transpose, Euclidean norm, and Frobenius norm of matrix $\mb{A}$ by $|\mb{A}|$, $\mb{A}^{\top}$, $\mb{A}^H$, $\|\mathbf{A}\|$, and $\|\mathbf{A}\|_{\mr{F}}$, respectively. The symbols $\sim$ and $\propto$ represent ``distributed according to'' and ``proportional to,'' respectively. The element in the $i^{\mathrm{th}}$ row and $j^{\mathrm{th}}$ column of matrix $\mathbf{X}$ is denoted as $[\mathbf{X}]_{ij}$. We use $[\mb{X}]_{:m}$ to indicate the $m^{\mr{th}}$ column of $\mb{X}$. Additional notations include $\mr{Tr}(\cdot)$ for the trace function, $\mr{exp}\{\cdot\}$ for the exponential function, and $\mr{sign}(\cdot)$ for the signum
function. %, and $\log_{10}(\cdot)$ for the logarithm with base $10$. 
The identity matrix of size $M$ is represented by $\mathbf{I}_M$. The complex conjugate of $x$ is indicated as $x^*$, and the real and imaginary parts of $x$ are denoted by $\Re\{x\}$ and $\Im\{x\}$, respectively. The probability density function (PDF) of a length-$K$ complex-valued random vector $\mathbf{x}\sim\mathcal{CN}(\mathbf{m}, \mathbf{C})$ is given by:
$\mc{CN}(\mb{x};\mathbf{m}, \mb{C})= \frac{1}{\pi^K |\mathbf{C}|} \exp\left\{- (\mathbf{x} - \mathbf{m})^H \mathbf{C}^{-1} (\mathbf{x} - \mathbf{m}) \right\}.
$ The expected value and variance of $x$ under the distribution $p(x)$ are denoted by $\mathbb{E}_{p(x)}[x]$ and $\mathrm{Var}_{p(x)}[x]$, respectively. We use $\langle x \rangle$, $\langle |x|^2 \rangle$, and $\tau^x$ to represent the mean, second moment, and variance of $x$ under a variational distribution $q(x)$. We denote the indicator function by $\mathbbm{1}(\cdot)$, which takes the value of one if the given condition is true and zero otherwise.

	%%%%%%%%%%%%%%%%%%%%%%%%%%%%%%%%%%%%%%%%%%%%%%%%%%%
	
	\section{System Model}
        \label{Sec:system_model}
\begin{figure}[t]
\vspace{-4mm}
\centering
  \includegraphics[trim = 0mm 0mm 0mm 0mm, clip, scale=7, width=0.8\linewidth, draft=false]{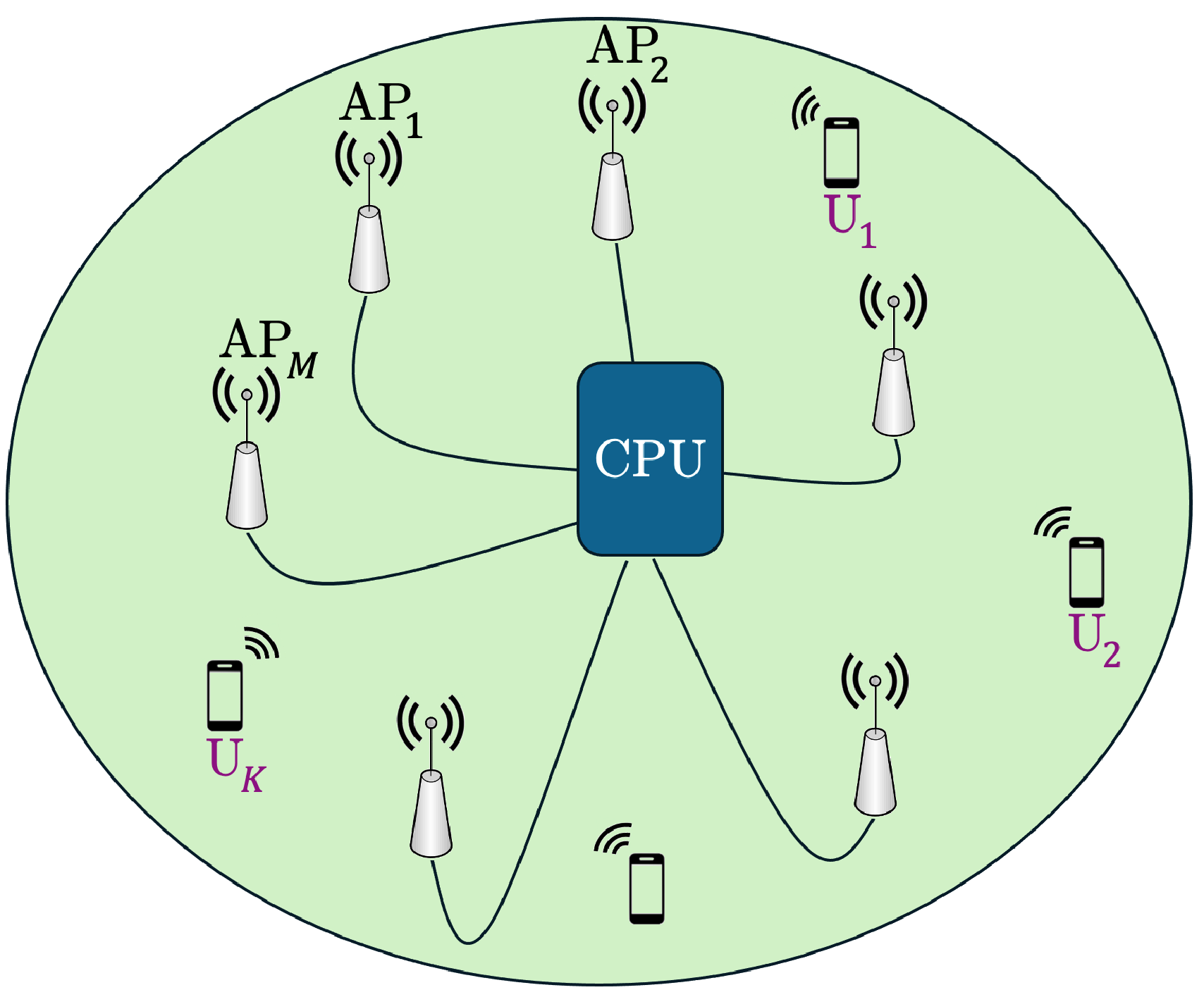}
  \vspace{-2mm}
  \caption{A CF-mMIMO network model with a CPU and multiple distributed
APs serving multiple single-antenna users.
  }\label{Fig:system_model}
  \vspace{-3mm}
\end{figure}

        In this study, we consider an uplink scenario in a CF-mMIMO network, depicted in Fig.~\ref{Fig:system_model}, where a CPU utilizes $L$ distributed APs to concurrently serve $K$ users. Each AP is equipped with $M$ antennas, while each user has a single antenna. Our focus is on addressing a JED problem within this network. To this end, we develop a method leveraging VB inference to approximate the true posterior distributions. This section begins by describing the channel model, provides a brief introduction to VB inference, and then presents the problem formulation.

\subsection{Channel Model}
We use $\mb{h}_{i,\ell} = [h_{i,\ell}^{[1]}, h_{i,\ell}^{[2]}, \ldots, h_{i,\ell}^{[M]}]^{\top} \in \mathbb{C}^{M\times 1}, 1 \leq i \leq K, 1 \leq \ell \leq L$, to denote the communication channel between the $i^{\mathrm{th}}$ user and the $\ell^{\mathrm{th}}$ AP, where $h_{i,\ell}^{[m]}, 1 \leq m \leq M,$  represents the channel between the $m^{\mathrm{th}}$ antenna element of the $\ell^{\mathrm{th}}$ AP and the $i^{\mathrm{th}}$ user. In this paper, we consider $1 \leq K \leq ML$
%\footnote{{\color{blue}While condition $M \leq K$ is not a fundamental requirement in CF-mMIMO, we adopt it here to highlight a key benefit of such systems compared to traditional mMIMO, where $M \geq K$ is essential to serve all users with a single AP.}} 
and assume communication over a coherence interval during which the channels remain constant. %We also consider %flat fading channels during the coherence time
%$M \leq K \leq ML$\footnote{{\color{blue}While condition $M \leq K$ is not a fundamental requirement in CF-mMIMO, we adopt it here to highlight a key benefit of such systems compared to traditional mMIMO, where $M \geq K$ is essential to serve all users with a single AP.}}, which is a typical scenario in CF-mMIMO networks.

To characterize $\mb{h}_{i,\ell}$, we assume that $\mb{h}_{i,\ell} \sim \mc{CN}(\mb{0}, \bs{\Sigma}_{i,\ell})$, where $\bs{\Sigma}_{i,\ell}$ is the channel covariance matrix, given by:
\begin{align}
    \bs{\Sigma}_{i,\ell} = \beta_{i,\ell}^{-1}\bar{\bs{\Sigma}}_{\ell},
\end{align}
where $\beta_{i,\ell}$ denotes the large-scale fading between the $i^{\mathrm{th}}$ user and the $\ell^{\mathrm{th}}$ AP, and $\bar{\bs{\Sigma}}_{\ell}$ is the spatial correlation between the antenna elements at the $\ell^{\mathrm{th}}$ AP. We assume that the channel covariance matrix $\boldsymbol{\Sigma}_{i,\ell}$ is known, which is a common assumption in the literature \cite{takahashi2022bayesian,nguyen2024variational}. We consider that the channels are independent across users, which leads to $\mathbb{E}[\mb{h}_{i,\ell}\mb{h}_{j,\ell}^H] = \mb{0}$ for $1 \leq \ell \leq L$, $1 \leq i, j \leq K$, and $i \neq j$. 

{\bf Received signal:} In the considered CF-mMIMO network, the received signal at duration $t$ at the $ m^{\mr{th}}$ antenna element of the $\ell^{\mr{th}}$ AP is denoted by $r_{m,\ell,t}$ and is modeled as:
\begin{eqnarray}
\label{system-model_signal_m}
    r_{m,\ell,t} = \sum_{i=1}^K h_{i,\ell}^{[m]} x_{i,t} + n_{m,\ell,t},
\end{eqnarray}
where $x_{i,t}$ represents the transmitted signal from the $i^{\mathrm{th}}$ user at duration $t$, while $n_{m,\ell,t} \sim \mathcal{CN}(0, N_0)$ is the i.i.d. AWGN at the $m^{\mathrm{th}}$ antenna element of the $\ell^{\mathrm{th}}$ AP at duration $t$. We then organize the received signals, channels, transmitted signals, and noise elements into vectors to formulate a linear model for the received signals at the $\ell^{\mathrm{th}}$ AP at duration $t$ as below:
\begin{eqnarray}
\label{system-model}
    \mb{r}_{\ell,t} = \mb{H}_{\ell}\mb{x}_t + \mb{n}_{\ell,t},
\end{eqnarray}
where $\mb{r}_{\ell,t} = [r_{1,\ell,t}, r_{2,\ell,t}, \ldots, r_{M,\ell,t}]^{\top} \in\mathbb{C}^{M\times 1 }$, $\mb{x}_t=[x_{1,t},x_{2,t},\ldots,x_{K,t}]^{\top}\in \mathbb{C}^{K\times 1 }$, $\mb{H}_{\ell}=\left[\mb{h}_{1,\ell},\mb{h}_{2,\ell},\ldots,\mb{h}_{K,\ell}\right]\in \mathbb{C}^{M\times K }$, and $\mb{n}_{\ell,t} = [n_{1,\ell,t}, n_{2,\ell,t}, \ldots, n_{M,\ell,t}]^{\top}\in\mathbb{C}^{M\times 1 }$. 

In this work, we examine an uplink CF-mMIMO network with the goal of jointly estimating $\mb{H}=\left[\mb{H}^{\top}_{1},\mb{H}^{\top}_{2},\ldots,\mb{H}^{\top}_{L}\right]^{\top} \in \mathbb{C}^{ML\times K}$ and detecting $\mb{x}_t$ from the received signal $\mb{r}_t = [\mb{r}^{\top}_{1,t}, \mb{r}^{\top}_{2,t}, \ldots, \mb{r}^{\top}_{L,t}] ^{\top} \in \mathbb{C}^{ML\times 1}$. To do this, we propose an approach relying on VB inference. Prior to exploring the methodology in-depth, we provide a concise introduction to the core principles of VB.

\subsection{Introduction to Variational Bayesian Inference} 
Variational Bayesian (VB) inference is a powerful statistical approach rooted in machine learning, designed to approximate the posterior distribution of latent variables efficiently. Consider the set of observed variables $\mb{r}$ and the set of $V$ latent variables $\mb{x}$. To detect $\mb{x}$, it is necessary to evaluate the posterior distribution $p(\mb{x}|\mb{r})$, which is often computationally intractable. To address this, VB approximates $p(\mb{x}|\mb{r})$ using a distribution $q(\mb{x})$ parameterized by variational variables, chosen from a predefined family $\mc{Q}$. The goal is to ensure that $q(\mb{x})$ is as close as possible to $p(\mb{x}|\mb{r})$. This is achieved by formulating an optimization problem that minimizes the KL divergence from $q(\mb{x})$ to $p(\mb{x}|\mb{r})$: \begin{equation} \label{eq:opt_KL} q^{\star}(\mb{x}) = \arg\min_{q(\mb{x}) \in \mc{Q}} \text{KL}\big(q(\mb{x}) | p(\mb{x}|\mb{r}) \big), \end{equation} where $q^{\star}(\mb{x})$ represents the optimal variational approximation, and the KL divergence is given by: 
\begin{equation} \text{KL}\big(q(\mb{x}) | p(\mb{x}|\mb{r}) \big) = \mathbb{E}_{q(\mb{x})} \big[\ln q(\mb{x}) \big] - \mathbb{E}_{q(\mb{x})} \big[\ln p(\mb{x}|\mb{r})\big]. 
\end{equation}

The KL divergence reaches its minimum when $q(\mb{x})$ matches the true posterior $p(\mb{x}|\mb{r})$. However, since deriving the exact posterior is usually infeasible, a practical approach involves restricting $q(\mb{x})$ to a simplified family of distributions. A common choice in the literature~\cite{Bishop-2006} is the mean-field variational family, where the latent variables are assumed to be independent, leading to $q(\mb{x}) = \prod_{i=1}^V q_i(x_i)$. Within this framework, the optimal solution for $q_i(x_i)$ is given by~[31] 
\begin{align} \label{eq:q_start_prop} 
q_i^{\star}(x_i) &\propto \exp\left\{\big\langle{\ln p(\mb{r},\mb{x})}\big\rangle\right\}, 
\end{align} 
where $\langle \cdot \rangle$ denotes the expectation with respect to all latent variables except $x_i$, using the current variational densities $q_{-i}(\mb{x}_{-i}) = \prod_{j=1, j\neq i}^{V} q_{j}(x_{j})$.

To solve the optimization problem in~\eqref{eq:opt_KL}, the Coordinate Ascent Variational Inference (CAVI) algorithm is commonly employed. This iterative method sequentially updates each $q_i^{\star}(x_i)$, as shown in \eqref{eq:q_start_prop}, while keeping the others fixed, thereby ensuring a monotonic improvement in the objective function in~\eqref{eq:opt_KL}. The CAVI algorithm is guaranteed to converge to a local optimum~\cite{Bishop-2006, Wainwright-2008}.

%==============================%

\subsection{Problem Formulation}
Taking into account the received signals at all APs results in the following system model at duration $t$:
\begin{eqnarray}
\label{system-model_stacked}
    \mb{r}_t = \mb{H}\mb{x}_t + \mb{n}_t,
\end{eqnarray}
where $\mb{n}_t = [\mb{n}^{\top}_{1,t}, \mb{n}^{\top}_{2,t}, \ldots, \mb{n}^{\top}_{L,t}]^{\top}\in \mathbb{C}^{ML\times 1}$.

In this study, we partition each coherence interval, consisting of $T$ symbols, into two phases: a pilot transmission phase of length $T_p$ symbols and a data transmission phase of length $T_d$ symbols, satisfying $T_p + T_d = T$. Here, we use $\mb{R}_p$ and $\mb{R}_d$ to denote the received signal matrices during the pilot and data transmission phases, respectively, which are given by:
\begin{align}
     \mb{R}_p &= \mb{H}\mb{X}_p + \mb{N}_p, \\ 
     \mb{R}_d &= \mb{H}\mb{X}_d + \mb{N}_d,
\end{align}
where $\mb{R}_p = [\mb{r}_{1}, \mb{r}_{2}, \ldots, \mb{r}_{T_p}] \in \mathbb{C}^{ML\times T_p}$, $\mb{X}_p = [\mb{x}_{1}, \mb{x}_{2}, \ldots, \mb{x}_{T_p}] \in \mathbb{C}^{K\times T_p}$, $\mb{N}_p = [\mb{n}_{1}, \mb{n}_{2}, \ldots, \mb{n}_{T_p}] \in \mathbb{C}^{ML\times T_p}$, $\mb{R}_d = [\mb{r}_{T_p + 1}, \mb{r}_{T_p + 2}, \ldots, \mb{r}_{T}] \in \mathbb{C}^{ML\times T_d}$, $\mb{X}_d = [\mb{x}_{T_p + 1}, \mb{x}_{T_p + 2}, \ldots, \mb{x}_{T}] \in \mathbb{C}^{K\times T_d}$, and $\mb{N}_d = [\mb{n}_{T_p + 1}, \mb{n}_{T_p + 2}, \ldots, \mb{n}_{T}] \in \mathbb{C}^{ML\times T_d}$.

% From now on, we use indices $p$ and $d$ to denote the parameters (variables) belonging to the pilot transmission phase and data transmission phase, respectively. 

In this paper, we use $\bs{\zeta}_p$ and $\bs{\zeta}_d$ to denote the residual inter-user interference during the pilot and data transmission phases, respectively. These terms capture uncertainty in the JED process, which includes noise, CE error, and data detection error, and are defined as follows:
\begin{align} 
\label{eq:zeta_p_paper} 
&\bs{\zeta}_p = \mb{R}_p - \mb{H}\mb{X}_p, \\
\label{eq:zeta_d_paper} 
&\bs{\zeta}_d = \mb{R}_d - \mb{H}\mb{X}_d.
\end{align}

We represent the estimated values of $\mb{H}$ and $\mb{X}_d$ by $\hat{\mb{H}} = \mb{H} + \mb{e}_{\rm{h}}$ and $\hat{\mb{X}}_d = \mb{X}_d + \mb{e}_{\rm{x}}$, respectively, where $\mb{e}_{\rm{h}}$  is the CE error, and $\mb{e}_{\rm{x}}$ denotes the data detection error. By substituting these estimated values into \eqref{eq:zeta_p_paper} and \eqref{eq:zeta_d_paper}, we obtain: 
\begin{align} \label{eq:zeta_p_error_paper} 
\bs{\zeta}_p &= \mb{R}_p - \hat{\mb{H}}\mb{X}_p = \mb{R}_p - (\mb{H} + \mb{e}_{\rm{h}})\mb{X}_p \\ \nonumber
&= \mb{N}_p - \mb{e}_{\rm{h}}\mb{X}_p,\\ 
\label{eq:zeta_d_error_paper} 
\bs{\zeta}_d &= \mb{R}_d - \hat{\mb{H}}\hat{\mb{X}}_d = \mb{R}_d - (\mb{H} + \mb{e}_{\rm{h}})(\mb{X}_d + \mb{e}_{\rm{x}}) \\ \nonumber
&= \mb{N}_d - \mb{H}\mb{e}_{\rm{x}} - \mb{e}_{\rm{h}}\mb{X}_d - \mb{e}_{\rm{h}}\mb{e}_{\rm{x}}. 
\end{align}

To effectively capture the uncertainty, we model $\bs{\zeta}_p$ and $\bs{\zeta}_d$ as i.i.d. zero-mean Gaussian random variables. We then define $\gamma_p$ and $\bs{\gamma}_d = [\gamma_{d,T_p+1}, \gamma_{d,T_p+2}, \ldots, \gamma_{d,T}]^{\top}$ to represent the precision of the combined effect of noise and CE error during the pilot transmission phase, and the precision of the combined effect of noise, CE error, and data detection error during the data transmission phase, respectively. %Since the underlying errors in \eqref{eq:zeta_p_error_paper} and \eqref{eq:zeta_d_error_paper} are unknown, we treat the precision parameters $\gamma_p$ and $\bs{\gamma}_d$ as unknown i.i.d. random variables and estimate them within the VB framework.
Since the underlying errors in \eqref{eq:zeta_p_error_paper} and \eqref{eq:zeta_d_error_paper} are assumed to be unknown Gaussian variables, we treat the precision parameters $\gamma_p$ and $\bs{\gamma}_d$ as unknown i.i.d. Gamma random variables and estimate them within the VB framework.

Notably, in the ideal scenario where both CE and data detection are error-free, the residual inter-user interference simplifies to the noise term alone. As a result, $\gamma_p$ and $\bs{\gamma}_d$ represent the noise precision. %In the next section, we will demonstrate the process of estimating $\gamma_p$ and $\bs{\gamma}_d$ using the EM technique. 

Based on \eqref{eq:q_start_prop}, in order to apply the VB method, it is necessary to compute the joint distribution, $p(\mb{R}_p, \mb{R}_d,  \mb{X}_d, \mb{H}, \gamma_p, \bs{\gamma}_d;  \mb{X}_p, \bs{\Sigma})$, which can be derived as follows:
\begin{align}
\label{eq:first_cond_prob}	
&p(\mb{R}_p, \mb{R}_d, \mb{X}_d, \mb{H}, \gamma_p, \bs{\gamma}_d; \mb{X}_p,\bs{\Sigma}) =p(\mb{R}_p | \mb{H}, \gamma_p; \mb{X}_p)\\ \nonumber
& \times  p(\mb{R}_d | \mb{X}_d, \mb{H}, \bs{\gamma}_d) p(\mb{H};\bs{\Sigma})p(\mb{X}_d)p(\gamma_p) p(\bs{\gamma}_d)\\ \nonumber
&=\left[ \prod_{t=1}^{T_p}p(\mb{r}_t|\mb{H},\gamma_p;\mb{x}_t) \right] \hspace{-2mm}\left[ \prod_{t=T_p + 1}^{T}p(\mb{r}_t|\mb{x}_t,\mb{H},\gamma_{d,t})p(\mb{x}_t) p(\gamma_{d,t})\right] \\ \nonumber
& \quad \quad \times p(\mb{H};\bs{\Sigma}) p(\gamma_p)\nonumber,
\end{align}
where $p(\mb{H};\bs{\Sigma}) = \prod_{i=1}^K \prod_{\ell=1}^Lp(\mb{h}_{i,\ell};\bs{\Sigma}_{i,\ell})$, $p(\mb{x}_t) = \prod_{i=1}^{K}p(x_{i,t})$, $\bs{\Sigma} = [\bs{\Sigma}_1, \bs{\Sigma}_2, \ldots, \bs{\Sigma}_K]$, and $\bs{\Sigma}_i = [\bs{\Sigma}_{i,1}, \bs{\Sigma}_{i,2}, \ldots, \bs{\Sigma}_{i,L}]$. 

In this study, our objective is to compute the Bayesian optimal estimates for both the data symbol $\mb{X}_d$ and the channel matrix $\mb{H}$. Achieving this requires the posterior distribution $p(\mb{X}_d, \mb{H}, \gamma_p, \bs{\gamma}_d| \mb{R}_p, \mb{R}_d, \mb{X}_p, \bs{\Sigma})$, which is often difficult to determine. Hence, we use the VB framework to approximate this posterior distribution, with detailed explanations provided in the subsequent section.
 
	%%%%%%%%%%%%%%%%%%%%%%%%%%%%%%%%%%%%%%%%%%%%%%%%%%%

	\section{VB Inference Framework For the JED Problem in CF-mMIMO}
        \label{Sec:VB_for_JED}
        % In this section, we outline the proposed VB-based method designed to tackle the JED problem in the uplink of a CF-mMIMO network. The goal is to compute the Bayesian optimal estimates for both the data symbols $\mb{X}_d$ and the channel matrix $\mb{H}$. Achieving this requires the posterior distribution $p(\mb{X}_d, \mb{H}, \gamma_p, \bs{\gamma}_d| \mb{R}_p, \mb{R}_d, \mb{X}_p, \bs{\Sigma})$, which is often difficult to determine. %Here, $\hat{\gamma}_p$ and $\hat{\bs{\gamma}}_d$ represent the estimated values of $\gamma_p$ and $\bs{\gamma}_d$, respectively. 
% To address this challenge, we adopt the mean-field variational distribution $q(\mb{X}_d, \mb{H}, \gamma_p, \bs{\gamma}_d)$ within the VB framework, as described in the following.
In this section, we propose a VB-based method to address the JED problem in the uplink of a CF-mMIMO network by approximating the intractable posterior distribution \( p(\mb{X}_d, \mb{H}, \gamma_p, \bs{\gamma}_d | \mb{R}_p, \mb{R}_d, \mb{X}_p, \bs{\Sigma}) \). To this end, we adopt a mean-field variational distribution \( q(\mb{X}_d, \mb{H}, \gamma_p, \bs{\gamma}_d) \) within the VB framework, as described in the following.
\begin{align} 
    \label{eq:mean_field}
&p\left(\mb{X}_d,\mb{H}, \gamma_p, \bs{\gamma}_d|\mb{R}_p, \mb{R}_d, \mb{X}_p, \bs{\Sigma} \right)
    \approx q(\mb{X}_d,\mb{H}, \gamma_p, \bs{\gamma}_d) \\
    &  \hspace{-1mm}= \left[\prod_{t=T_p + 1}^{T} \prod_{i=1}^K q_i(x_{i,t}) \right] \hspace{-2mm}\left[ \prod_{i=1}^K \prod_{\ell=1}^L 
    q(\mb{h}_{i,\ell})\right] \hspace{-2mm}\left[\prod_{t=T_p + 1}^{T} q(\gamma_{d,t}) \right]
    q(\gamma_p).\nonumber 
\end{align} 
% According to~\eqref{eq:q_start_prop}, obtaining the optimal variational densities in \eqref{eq:mean_field} requires the joint distribution {\color{blue}$p(\mb{R}_p, \mb{R}_d, \mb{X}_d, \mb{H}, \gamma_p, \bs{\gamma}_d; \mb{X}_p, \bs{\Sigma})$}, which can be expressed as
Based on~\eqref{eq:q_start_prop}, computing the optimal variational densities in \eqref{eq:mean_field} requires the joint distribution $p(\mb{R}_p, \mb{R}_d, \mb{X}_d, \mb{H}, \gamma_p, \bs{\gamma}_d; \mb{X}_p, \bs{\Sigma})$, given by:
\begin{align} 
\label{conditional}
    &p(\mb{R}_p, \mb{R}_d, \mb{X}_d,\mb{H}, \gamma_p, \bs{\gamma}_d; \mb{X}_p, \bs{\Sigma}) = \left[\prod_{t=1}^{T_p}p(\mb{r}_t| \mb{H}, \gamma_p; \mb{x}_t)\right] \nonumber \\
    &\quad \times \left[\prod_{t=T_p + 1}^{T} p(\mb{r}_t| \mb{x}_t,\mb{H},\gamma_{d,t})\right] \left[\prod_{t=T_p + 1}^{T} \prod_{i=1}^K p(x_{i,t})\right] \nonumber \\
    &\quad  \times \left[\prod_{i=1}^K \prod_{\ell =1} ^ L  p(\mb{h}_{i,\ell};\bs{\Sigma}_{i,\ell}) \right]\left[\prod_{t=T_p + 1}^{T} p(\gamma_{d,t})\right] p(\gamma_p).
\end{align}

To use~\eqref{conditional} within the VB framework, we need to specify the prior distributions for $p(x_{i,t})$ and $p(\gamma_{d,t})$ for $T_p+1 \leq t \leq T$, $p(\gamma_p)$, and $p(\mb{h}_{i,\ell}; \bs{\Sigma}_{i,\ell})$. We assume the prior distribution of $x_{i,t}$ as $p(x_{i,t}) = \sum_{a \in \mc{S}} p_a \delta(x_{i,t} - a)$, where $p_a$ represents the probability of the constellation point $a \in \mc{S}$, and $\mc{S}$ denotes the signal constellation. Moreover, we assume $\gamma_p \sim \Gamma(a_p, b_p)$, $\gamma_{d,t} \sim \Gamma(a_{d,t}, b_{d,t})$, and $p(\mb{h}_{i,\ell}; \bs{\Sigma}_{i,\ell}) = \mc{CN}(\mb{h}_{i,\ell};\bs{0}, \bs{\Sigma}_{i,\ell})$. %where $\hat{\mb{h}}_{i,\ell}$ and $\hat{\bs{\Sigma}}_{i,\ell}$ are the corresponding mean and covariance, respectively. 
%In Table~\ref{table:prior_dist_online}, we summarize the prior distribution assumptions for the desired unknown random variables $p(\mb{h}_{i,\ell})$, $p(\gamma_p)$, $p(\gamma_{d,t})$, and $p(x_{i,t})$.

% \begin{table} [ht]
% \caption{The prior distribution assumptions for the unknown random variables in the proposed strategy relying on the VB framework
% }
% \centering
% {
% \begin{tabular}{|c|c|}
% \hline
% Probability & Prior Distribution Assumption\\%[0.5ex]
% \hline \hline
% \rule{0pt}{10pt}
% $p(\mb{h}_{i,\ell})$ &  $\mc{CN}\big(\bs{0},\bs{\Sigma}_{i,\ell}\big)$\\

% $p(\gamma_p)$ & $\Gamma(a_{0,p},b_{0,p})$\\

% $p(\gamma_{d,t})$ & $\Gamma(a_{0,d},b_{0,d})$\\

% $p(x_{i,t}) $ & $\sum_{a\in \mc{S}} p_a\delta(x_{i,t} - a)$\\

% \hline
% \end{tabular}
% }
% \label{table:prior_dist_online}
% \end{table}

JED processing is performed at the CPU and can operate on either unquantized or quantized signals forwarded by the APs. While forwarding unquantized signals requires high fronthaul bandwidth, transmitting quantized signals from the APs to the CPU reduces the bandwidth demand. To analyze these situations, we consider the following three scenarios.

\subsection{ Scenario 1 -- Perfect Fronthaul Link}
\label{subsect:Case1_Full_Center}

In the PFL scenario, we assume that all APs forward their unquantized received signals to the CPU. We then use the CAVI algorithm to find the local optimal solutions: $q^{\star}(\mb{h}_{i,\ell})$, $q^{\star}(x_{i,t})$, $q^{\star}(\gamma_{d,t})$, and $q^{\star}(\gamma_p)$. % by optimizing one latent variable at a time while the others are held fixed. %Furthermore, as defining a prior distribution for noise precision can be challenging, we leverage the EM method to estimate $\gamma_p$ and $\bs{\gamma}_d$ within the CAVI algorithm, which relies only on the observed received signals from APs. 
In the following subsections, we describe the update procedure for each of these variables.

\subsubsection{ Updating $\mb{h}_{i,\ell}$} By evaluating the expectation of~\eqref{conditional} over all latent variables except $\mb{h}_{i,\ell}$, we derive the variational distribution $q(\mb{h}_{i,\ell})$ as shown in~\eqref{q-h-i} (at the top of the next page). 
\begin{figure*}
    \begin{align} 
    \label{q-h-i}
        q(\mb{h}_{i,\ell}) &\propto \mr{exp}\left\{ \bblr{\ln \Big[\prod_{t=1}^{T_p}p(\mb{r}_t| \mb{H},\gamma_p; \mb{x}_t)\Big] + \ln \Big[\prod_{t=T_p + 1}^{T}p(\mb{r}_t| \mb{x}_t, \mb{H},\gamma_{d,t})\Big] + \ln p(\mb{h}_{i,\ell};\bs{\Sigma}_{i,\ell})}\right\}  \nonumber \\
        &\propto \mr{exp}\Big\{- \blr{\gamma_p \sum_{t=1}^{T_p} \big\|\mb{r}_t-\mb{H}\mb{x}_t\big\|^2} - \blr{ \sum_{t=T_p + 1}^{T} \gamma_{d,t} \big\|\mb{r}_t-\mb{H}\mb{x}_t\big\|^2} - \blr{ \mb{h}_{i,\ell}^H\bs{\Sigma}^{-1}_{i,\ell}\mb{h}_{i,\ell}}  \Big\} \nonumber  \\
        &\propto \mr{exp}\bigg\{-\mb{h}_{i,\ell}^H\Big[ \big (\lr{\gamma_p}\sum_{t = 1}^{T_p}|x_{i,t}|^2 + \sum_{t= T_p + 1}^{T}\lr{\gamma_{d,t}}\lr{|x_{i,t}|^2} \big) \mb{I}_M + \bs{\Sigma}^{-1}_{i,\ell}\Big]\mb{h}_{i,\ell}  \nonumber \\
        &\quad\quad + 2\,\Re\Big\{\lr{\gamma_p}\mb{h}_{i,\ell}^H \sum_{t=1}^{T_p}\bigg(\mb{r}_{\ell,t} - \sum_{j=1, j\neq i}^K \lr{\mb{h}_{j,\ell}} x_{j,t} + \sum_{\ell^{\prime} = 1, \ell^{\prime} \neq \ell}^{L} \left(\mb{r}_{\ell^{\prime},t} - \lr{\mb{H}_{\ell^{\prime}}} \mb{x}_t\right)\bigg)x_{i,t}^* \nonumber \\
        &\quad\quad + \mb{h}_{i,\ell}^H \sum_{t = T_p + 1}^{T} \lr{\gamma_{d,t}} \bigg(\mb{r}_{\ell,t} - \sum_{j=1, j\neq i}^K \lr{\mb{h}_{j,\ell}} \lr{x_{j,t}} + \sum_{\ell^{\prime} = 1, \ell^{\prime} \neq \ell}^{L} \left(\mb{r}_{\ell^{\prime},t} - \lr{\mb{H}_{\ell^{\prime}}} \lr{\mb{x}_t}\right)\bigg)\lr{x_{i,t}^*}\Big\}\bigg\}.
    \end{align}
    \hrulefill
    \vspace{-5mm}
\end{figure*}
Given that $\mb{h}_{i,\ell}$ is assumed to follow a Gaussian prior distribution, it is reasonable to guess that $q(\mb{h}_{i,\ell})$ will also be Gaussian. Our derivations in~\eqref{q-h-i} confirm this point. Specifically, $q(\mb{h}_{i,\ell})$ is Gaussian with the following covariance matrix and mean: 
\begin{align}
\label{Sigma-h}
    \hat{\bs{\Sigma}}_{i,\ell} &= \Big[\Big(\lr{\gamma_p}\sum_{t = 1}^{T_p}|x_{i,t}|^2 + \sum_{t= T_p + 1}^{T}\lr{\gamma_{d,t}}\lr{|x_{i,t}|^2} \Big) \mb{I}_{M} \\ \nonumber
    & \qquad \qquad+ \bs{\Sigma}^{-1}_{i,\ell}\Big]^{-1}, 
\end{align}
\begin{align}
\label{mean-h}
    \lr{\mb{h}_{i,\ell}} &=\hat{\bs{\Sigma}}_{i,\ell}\Big[\lr{\gamma_p}\sum_{t=1}^{T_p}\Big(\mb{r}_{\ell,t} - \sum_{j=1, j\neq i}^K \lr{\mb{h}_{j,\ell}} x_{j,t} \Big)x_{i,t}^*\\ \nonumber
    &+ \lr{\gamma_p}\sum_{t=1}^{T_p}\Big(\sum_{\ell^{\prime} = 1, \ell^{\prime} \neq \ell}^{L} \left(\mb{r}_{\ell^{\prime},t} - \lr{\mb{H}_{\ell^{\prime}}} \mb{x}_t\right)\Big) x_{i,t}^*  \\ \nonumber
    &+ \sum_{t=T_p + 1}^{T}\lr{\gamma_{d,t}}\Big(\mb{r}_{\ell,t} - \sum_{j=1, j\neq i}^K \lr{\mb{h}_{j,\ell}} \lr{x_{j,t}} \Big)\lr{x_{i,t}^*}\\ \nonumber
    &+ \sum_{t= T_p + 1}^{T}\lr{\gamma_{d,t}}\Big(\sum_{\ell^{\prime} = 1, \ell^{\prime} \neq \ell}^{L} \left(\mb{r}_{\ell^{\prime},t} - \lr{\mb{H}_{\ell^{\prime}}} \lr{\mb{x}_t}\right)\Big)\lr{x_{i,t}^*} \Big].
\end{align}
%------------------------------------------

%------------------------------------------

We use the following lemma to compute the variational posterior mean of $x_{i,t}$, $\gamma_p$, and $\bs{\gamma}_{d}$.
\begin{lemma}
\label{theorem-2}
\cite{nguyen2024variational} Let $\mb{y} \in \mathbb{C}^{m\times 1}$, $\mb{A}\in \mathbb{C}^{m\times n}$, and $\mb{x}\in \mathbb{C}^{n\times 1}$ be three independent random matrices (vectors) with respect to a variational density $q_{\mb{y},\mb{A},\mb{x}}(\mb{y},\mb{A},\mb{x}) = q_{\mb{y}}(\mb{y}) q_{\mb{A}}(\mb{A})q_{\mb{x}}(\mb{x})$. Suppose $\mb{A}$ is column-wise independent and $\lr{\mb{a}_i}$ and $\bs{\Sigma}_{\mb{a}_i}$ are the variational mean and covariance matrix of the $i^{th}$ column of $\mb{A}$. Let $\lr{\mb{y}}$ and $\bs{\Sigma}_{\mb{y}}$ be the variational mean and covariance matrix of $\mb{y}$ and $\lr{\mb{x}}$ and $\bs{\Sigma}_{\mb{x}}$ be the variational mean and covariance matrix of $\mb{x}$. Consider $\mb{F}$ is an arbitrary Hermitian matrix. Here, $\blr{(\mb{y}-\mb{A}\mb{x})^H \mb{F} (\mb{y}-\mb{A}\mb{x})}$ with respect to $q_{\mb{y},\mb{A},\mb{x}}(\mb{y},\mb{A},\mb{x})$, is given by:
\begin{align}\label{f-ABC}
    &\blr{(\mb{y}-\mb{A}\mb{x})^H \mb{F} (\mb{y}-\mb{A}\mb{x})} \nonumber \\
    &= \big(\lr{\mb{y}}-\lr{\mb{A}}\lr{\mb{x}}\big)^H \mb{F} \big(\lr{\mb{y}}-\lr{\mb{A}}\lr{\mb{x}}\big) +  \lr{\mb{x}}^H\mb{B}\lr{\mb{x}}   \nonumber \\
    &+\tr\big\{\mb{F}\bs{\Sigma}_{\mb{y}}\big\}+ \tr\big\{\bs{\Sigma}_{\mb{x}}\mb{B}\big\}+\tr\big\{\bs{\Sigma}_{\mb{x}}\lr{\mb{A}^H}\mb{F}\lr{\mb{A}} \big\},
\end{align}
where $\mb{B} = \mr{diag}\big(\tr\{\mb{F}\bs{\Sigma_{\mb{a}_1}}\},\ldots, \tr\{\mb{F}\bs{\Sigma_{\mb{a}_n}}\}\big)$. 
\end{lemma}
\begin{IEEEproof}
The proof of this lemma is provided in \cite{nguyen2024variational}.
% The proof of this lemma can be found in \cite{nguyen2024variational}. For the sake of brevity, we refer the reader to \cite{nguyen2024variational} for detailed proof.
\end{IEEEproof}
	
% \begin{corollary} \label{corol_1}
% Given $\mb{x}$ is deterministic, $\blr{\|\mb{y}-\mb{A}\mb{x}\|^2}$ is simplified to:
%     \begin{eqnarray}\label{f-A}
%         \blr{\|\mb{y}-\mb{A}\mb{x}\|^2} = \|\mb{y}-\lr{\mb{A}}{\mb{x}}\|^2 
%         + \sum_{i=1}^n |x_i|^2 \tr\{\hat{\bs{\Sigma}}_{\mb{a}_i}\}.
%     \end{eqnarray}
% \end{corollary}
% \begin{IEEEproof}
%     We derive~\eqref{f-A} by setting $\hat{\bs{\Sigma}}_{\mb{x}} = \mb{0}$ and  ${\mb{x}}^H\mb{D}\mb{x}  =   \sum_{i=1}^n |x_i|^2 \tr\{\hat{\bs{\Sigma}}_{\mb{a}_i}\}$ in Lemma~\ref{theorem-2}.
% \end{IEEEproof}

%-----------------------------------------

\subsubsection{ Updating $x_{i,t}$} We apply this update exclusively during the data transmission phase. We compute the expectation of \eqref{conditional} over all latent variables, excluding $x_{i,t}$, to derive the variational distribution $q_i(x_{i,t})$, as shown in \eqref{x-t-i} on the following page,
\begin{figure*}
\begin{align} 
    q_i(x_{i,t}) &\propto \mr{exp}\Big\{\blr{\ln p(\mb{r}_t|\mb{x}_t,\mb{H},\bs{\gamma}_d) + \ln p(x_{i,t})}\Big\}\nonumber \\
    &\propto p(x_{i,t})\, \mr{exp}\left\{- \lr{\gamma_{d,t}} \bblr{\bigg\|\mb{r}_t-x_{i,t}\mb{h}_{i} - \sum_{j=1, j\neq i}^K x_{j,t}\mb{h}_{j}\bigg\|^2}\right\} \nonumber\\
    &\propto p(x_{i,t})\,\mr{exp}\left\{- \lr{\gamma_{d,t}} \bigg[ \blr{\|\mb{h}_{i}\|^2}|x_{i,t}|^2 - 2 \,\Re\bigg\{\!\blr{\mb{h}_{i}^H}\bigg(\mb{r}_t - \sum_{j=1, j\neq i}^K\!\lr{x_{j,t}}\blr{\mb{h}_{j}}  \bigg)x_{i,t}^*\bigg\}  \bigg]\right\} \nonumber \\
    &\propto  p(x_{i,t})\,\mr{exp}\left\{-\lr{\gamma_{d,t}} \blr{\|\mb{h}_{i}\|^2} |x_{i,t}-z_{i,t}|^2\right\}, \label{x-t-i}
\end{align}
\hrulefill
\vspace{-5mm}
\end{figure*}
where 
\begin{align}
    \mb{h}_{i} = [\mb{h}^{\top}_{i,1},\mb{h}^{\top}_{i,2},\ldots,\mb{h}^{\top}_{i,L}]^{\top},
\end{align}
and
\begin{align}
\label{eq:z_i,t}
z_{i,t}\triangleq\frac{\lr{\mb{h}_{i}^H}}{\blr{\|\mb{h}_{i}\|^2}}\Bigg(\mb{r}_t - \sum_{j=1, j\neq i}^K \lr{\mb{h}_{j}}\lr{x_{j,t}}\Bigg),
\end{align}
serves as a linear approximation for $x_{i,t}$. To evaluate $\big\langle\|\mb{h}_{i}\|^2\big\rangle$ in~\eqref{eq:z_i,t}, we utilize Lemma~\ref{theorem-2}, giving:
\begin{align}
    \big\langle\|\mb{h}_{i}\|^2\big\rangle = \|\langle\mb{h}_{i}\rangle\|^2 + \tr\{\hat{\bs{\Sigma}}_{i}\},
\end{align}
where $\hat{\bs{\Sigma}}_i = [\hat{\bs{\Sigma}}_{i,1}, \hat{\bs{\Sigma}}_{i,2}, \ldots, \hat{\bs{\Sigma}}_{i,L}]$. Since the prior distribution $p(x_{i,t})$ is discrete, the variational distribution $q_i(x_{i,t})$ is also discrete. Hence, normalization is required such that
\begin{align} 
q_i(a) = \frac{p_a\,\mr{exp}\big\{\! -\!\lr{\gamma_{d,t}} \blr{\|\mb{h}_{i}\|^2}|a-z_{i,t}|^2 \big\}}
{\sum_{b\in\mc{S}} p_b\,\mr{exp}\big\{\! -\!\lr{\gamma_{d,t}} \blr{\|\mb{h}_{i}\|^2}|a-z_{i,t}|^2 \big\}},\forall a\in\mc{S}.
\end{align} 

As a result, the expected value and variance of $x_{i,t}$ under the variational distribution are given by:
\begin{align}
\label{eq:mean_x_i_t}
	\lr{x_{i,t}} &= \sum_{a\in\mc{S}} a q_i(a),\\
\label{eq:var_x_i_t}
 \tau^x_{i,t} &= \sum_{a\in\mc{S}} |a|^2 q_i(a) - |\lr{x_{i,t}}|^2. 
\end{align}

%--------------------------------	

\subsubsection{Updating $\gamma_p$} Here, we determine the variational distribution $q(\gamma_p)$ by taking the expectation of \eqref{conditional} with respect to all latent variables except $\gamma_p$, leading to the following expression:
\begin{align} 
\label{eq:q(gamma_p)_ini}
    q(\gamma_p) & \propto\mr{exp}\Big\{\big\langle\ln \big[\prod_{t=1}^{T_p}p(\mb{r}_t|\mb{H},\gamma_p; \mb{x}_t) \big] + \ln p(\gamma_p)\big\rangle\Big\}  \nonumber \\
    &\propto \mr{exp}\Big\{M L T_p\ln \gamma_p - \gamma_p \sum_{t=1}^{T_p}\blr{\|\mb{r}_t-\mb{H}\mb{x}_t\|^2}\nonumber \\ & \quad \quad \quad \quad+ (a_p - 1) \ln \gamma_p - b_p \gamma_p \Big\}.
\end{align}
Based on \eqref{eq:q(gamma_p)_ini}, $q(\gamma_p)$ is Gamma distribution with mean
\begin{align} 
\label{eq:q(gamma_p)}
    \lr{\gamma_p} = \frac{ a_p + M L T_p}{b_p + \sum_{t=1}^{T_p}\blr{\|\mb{r}_t-\mb{H}\mb{x}_t\|^2}},
\end{align}
where $
\blr{\|\mb{r}_t-\mb{H}\mb{x}_t\|^2} = \|\mb{r}_t-\lr{\mb{H}}\mb{x}_t\|^2 + \sum\limits_{i=1}^K |x_{i,t}|^2 \tr\{\hat{\bs{\Sigma}}_{i}\}$ using Lemma \ref{theorem-2}.

\subsubsection{ Updating $\gamma_{d,t}$} Similar to the VB process for $\gamma_p$, we have:
\begin{align} 
\label{eq:q(gamma_d)_ini}
    q(\gamma_{d,t}) &\propto \mr{exp}\Big\{\big\langle\ln p(\mb{r}_t|\mb{x}_t,\mb{H},\gamma_{d,t}) + \ln p(\gamma_{d,t})\big\rangle\Big\}  \nonumber \\
    & \propto \mr{exp} \Big\{M L \ln \gamma_{d,t} -  \blr{\|\mb{r}_t-\mb{H}\mb{x}_t\|^2} \gamma_{d,t} \nonumber \\
    & \quad \quad \quad \quad + (a_{d,t}-1) \ln \gamma_{d,t} - b_{d,t}\gamma_{d,t}\Big\},
\end{align}
which results in $q(\gamma_{d,t})$ being Gamma distribution with the following mean:
\begin{align} 
\label{eq:q(gamma_d)}
    \lr{\gamma_{d,t}} = \frac{a_{d,t} + ML}{b_{d,t} + \blr{\|\mb{r}_t-\mb{H}\mb{x}_t\|^2}},
\end{align}
where $
\blr{\|\mb{r}_t-\mb{H}\mb{x}_t\|^2} = \|\mb{r}_t-\lr{\mb{H}}\lr{\mb{x}_t}\|^2 + \sum\limits_{i=1}^K\big[  \tau^x_{i,t}\|\lr{\mb{h}_{i}}\|^2 + \lr{|x_{i,t}|^2}\tr\{\hat{\bs{\Sigma}}_{i}\} \big]$ using Lemma \ref{theorem-2}.

If we set $a_p = b_p = a_{d,t} = b_{d,t} = 0$ in \eqref{eq:q(gamma_p)} and \eqref{eq:q(gamma_d)}, the expectations $\langle \gamma_p \rangle$ and $\langle \gamma_{d,t} \rangle$ become equivalent to the estimated values of the precision of the residual inter-user interference described in \cite{nguyen2024variational} using the EM technique. Hence, the VB-EM method can be seen as a special case of our proposed VB method, which employs Gamma priors for the residual terms.

It is essential to note that, in our VB framework, $\gamma_p$ is assumed to be a scalar due to the single-shot nature of the CE process over the entire pilot block, whereas $\bs{\gamma}_d$ is modeled as a vector to better capture the dynamic and iterative nature of the JED process during the data transmission phase.

% As discussed earlier, in the PFL scenario, the CPU needs to receive unquantized signals from all APs, % during both the pilot and data transmission phases,
% typically requiring $10$-bit quantizers to have nearly continuous signals \cite{takahashi2022bayesian}. This results in the transmission of $10ML(T_p + T_d)$ bits as fronthaul signaling overhead, % between the APs and the CPU, 
% demanding significant bandwidth. However, the limited bandwidth of the fronthaul links makes this approach impractical. Therefore, it becomes crucial to reduce the signaling between the APs and the CPU. To tackle this challenge, we introduce the Q-E and E-Q scenarios, which are discussed in the following subsections.

As discussed earlier, the PFL scenario requires APs to forward unquantized signals to the CPU, which typically demands high-resolution quantizers to approximate continuous signals \cite{takahashi2022bayesian}. However, this results in a high fronthaul bandwidth requirement, making the approach impractical in real-world systems. To address this challenge, we introduce the Q-E and E-Q scenarios, explained in the following subsections.

\subsection{Scenario 2 -- Quantization-and-Estimation}
\label{subsect:Q&E}

In this case, each AP quantizes its received signal to $b$ bits and then forwards it to the CPU. Then, the CPU employs the quantized received signals to perform JED. Here, $\mb{y}_{\ell,t}$ is the quantized received signal vector at the $\ell^{\mr{th}}$ AP at duration $t$, which is given by:
\begin{align}
    &\Re \{\mb{y}_{\ell,t}\}= \mc{Q}_{b}(\Re\{\mb{r}_{\ell,t}\}), \quad \Im \{\mb{y}_{\ell,t}\}= \mc{Q}_{b}(\Im\{\mb{r}_{\ell,t}\}),
\end{align}
where $\mc{Q}_{b}(\cdot)$ represents a quantization operator with $b$ bits, utilizing a uniform scalar quantization approach. It operates based on a set of $2^{b}-1$ thresholds, denoted as $\{d_1, d_2, \dots, d_{2^{b}-1}\}$. For convenience, we define the thresholds to satisfy $-\infty = d_0 < d_1 < d_2 < \dots < d_{2^{b}-1} < d_{2^{b}} = \infty$. The step size for quantization, denoted as $\Delta$, is used to express the quantization thresholds as $d_i = (-2^{b-1} + i)\Delta$ for $i = 1, 2, \dots, 2^{b}-1$. This results in a quantized output $q$ as follows:
\begin{eqnarray}
    q= Q_{b}(r)&=& \left\{ \begin{array}{ll}
        d_i-\Delta/2, &  r \in (d_{i-1}, d_i] \\
        0.5(2^{b}-1)\Delta, & r \in (d_{2^{b}-1},d_{2^{b}}].\\
    \end{array} \right. 
\end{eqnarray}

Further, we use $q^{\mr{low}} = d_{i-1}$ and $q^{\mr{up}} = d_i$ to represent the lower and upper bounds of the quantization bin that contains $q$. Then, in order to apply the VB framework, we need to formulate the joint probability as \eqref{eq:conditional_y_Q&E}, shown at the top of the next page,
\begin{figure*}
\begin{align}
\label{eq:conditional_y_Q&E}
    &p(\mb{Y}_p, \mb{Y}_d, \mb{R}_p, \mb{R}_d, \mb{X}_d, \mb{H}, \gamma_p, \bs{\gamma_d};\mb{X}_p, \bs{\Sigma}) \nonumber \\ 
    & \quad \quad  = \left[\prod_{t=1}^{T_p}p(\mb{y}_t|\mb{r}_t) p(\mb{r}_t|\mb{H}, \gamma_p;\mb{x}_t) p(\gamma_p)\right] \left[\prod_{t=T_p + 1}^{T}p(\mb{y}_t|\mb{r}_t) p(\mb{r}_t|\mb{x}_t, \mb{H}, \gamma_{d,t}) p(\mb{x}_t) p(\gamma_{d,t})\right] p(\mb{H}; \bs{\Sigma}) \nonumber \\ 
    & \quad \quad  = \left[\prod_{t=1}^{T_p}\prod_{m=1}^M\prod_{\ell=1}^L p(y_{m,\ell,t}|r_{m,\ell,t})p(r_{m,\ell,t}\big|[\mb{H}{_\ell}]_{:m}, \gamma_p; \mb{x}_t) \right] \left[\prod_{t=T_p + 1}^{T}  \prod_{m=1}^M\prod_{\ell=1}^L p(y_{m,\ell,t}|r_{m,\ell,t})p(r_{m,\ell,t}\big|\mb{x}_t, [\mb{H}{_\ell}]_{:m}, \gamma_{d,t})  \right]\nonumber \\ 
    & \quad \quad  \times \left[\prod_{t=T_p + 1}^{T}\prod_{i=1}^{K} p(x_{i,t}) \right]\left[\prod_{i=1}^{K}  \prod_{\ell=1}^{L}  p(\mb{h}_{i,\ell}; \bs{\Sigma}_{i,\ell}) \right]\left[\prod_{t=T_p + 1}^{T} p(\gamma_{d,t}) \right] p(\gamma_p),
\end{align}
\hrulefill
\vspace{-4mm}
\end{figure*}
where $\mb{y}_t$ is the stacked vector of $\mb{y}_{\ell,t}$ from all APs at duration $t$, $\mb{Y}_p = [\mb{y}_{1}, \mb{y}_{2}, \ldots, \mb{y}_{T_p}] \in \mathbb{C}^{ML\times T_p}$, $\mb{Y}_d = [\mb{y}_{T_p+1}, \mb{y}_{T_p+2}, \ldots, \mb{y}_{T}] \in \mathbb{C}^{ML\times T_d}$, $p(y_{m, \ell,t}|r_{m,\ell,t}) = \mathbbm{1}(r_{m,\ell,t} \in [y_{m,\ell,t}^{\mr{low}}, y_{m,\ell,t}^{\mr{up}}])$, $p(\mb{r}_t|\mb{H}, \gamma_p; \mb{x}_t) = \mc{CN}(\mb{r}_t;\mb{H}\mb{x}_t, \gamma_p^{-1}\mb{I}_{ML})$, and $p(\mb{r}_t|\mb{x}_t, \mb{H}, \gamma_{d,t}) = \mc{CN}(\mb{r}_t;\mb{H}\mb{x}_t, \gamma_{d,t}^{-1}\mb{I}_{ML})$. Here, our goal is to use the VB framework to get the variational distribution $q(\mb{R}_p, \mb{R}_d,\mb{X}_d,\mb{H}, \gamma_p, \bs{\gamma}_d)$ to approximate the posteriori distribution $p(\mb{R}_p, \mb{R}_d,\mb{X}_d,\mb{H}, \gamma_p, \bs{\gamma}_{d}| \mb{Y}_p, \mb{Y}_d; \mb{X}_p, \bs{\Sigma})$ as:
\begin{align}
    & p(\mb{R}_p, \mb{R}_d,\mb{X}_d,\mb{H}, \gamma_p, \bs{\gamma}_{d}| \mb{Y}_p, \mb{Y}_d; \mb{X}_p, \bs{\Sigma}) \\  \nonumber 
    &\approx q(\mb{R}_p, \mb{R}_d,\mb{X}_d,\mb{H}, \gamma_p, \bs{\gamma}_{d}) = \left[\prod_{t=1}^{T} q(\mb{r}_t)\right] \left[\prod_{i=1}^K \prod_{\ell=1}^L 
    q(\mb{h}_{i,\ell})\right] \\ \nonumber 
    & \times \left[ \prod_{t=T_p + 1}^{T} \prod_{i=1}^K q_i(x_{i,t})\right] \left[ \prod_{t=T_p + 1}^{T} q(\gamma_{d,t})\right]q(\gamma_p).
\end{align}

We then iteratively apply the CAVI algorithm to obtain the local optimal solutions $q^{\star}(\mb{r}_t)$, $q^{\star}(\mb{h}_{i,\ell})$, $q^{\star}(x_{i,t})$, $q^{\star}(\gamma_p)$, and $q^{\star}(\gamma_{d,t})$. The derivations for these variational distributions, except for $q^{\star}(\mb{r}_t)$, follow directly from the procedures established for the PFL scenario in Section \ref{subsect:Case1_Full_Center}. Therefore, we focus on detailing the steps for obtaining $q^{\star}(\mb{r}_t)$ given the quantized received signal $\mb{y}_t$ in the following part. 

\subsubsection{Updating $\mb{r}_t$} This update contains two parts:

{\bf Pilot Transmission Phase:} By computing the expectation of~\eqref{eq:conditional_y_Q&E} for all latent variables except for $\mb{r}_t$ when $1 \leq t \leq T_p$, we express the variational distribution $q(\mb{r}_t)$ as:
\begin{align}
\label{eq:q_r_p}
    q(\mb{r}_t) & \propto \exp \left \{ \lr{\ln p(\mb{y}_t|\mb{r}_t) + \ln p(\mb{r}_t|\mb{H}, \gamma_p; \mb{x}_t)} \right \} \\ \nonumber
    & \propto \mathbbm{1} (\mb{r}_t \in [\mb{y}_t^{\mr{low}}, \mb{y}_t^{\mr{up}}]) \times \exp\left \{ -\lr{\gamma_p} \| \mb{r}_t - \lr{\mb{H}} \mb{x}_t\|^2\right \}.
\end{align}

Notice the variational distribution $q(\mb{r}_t)$ is inherently separable as $\prod_{m=1}^M \prod_{\ell=1}^L q(r_{m,\ell,t})$ and the variational distribution
$q(r_{m,\ell,t})$ is the complex Gaussian distribution obtained
from bounding $r_{m,\ell,t} \sim \mc{CN}(\lr{[\mb{H}_{\ell}]_{:m}} \mb{x}_t, \gamma_p^{-1})$ to the interval $(y_{m,\ell}^{\mr{low}}, y_{m,\ell}^{\mr{up}})$. 

{\bf Data Transmission Phase:} Similar to the pilot transmission phase, we have:
\begin{align}
\label{eq:q_r_d}
    q(\mb{r}_t) & \propto \exp \left \{ \lr{\ln p(\mb{y}_t|\mb{r}_t) + \ln p(\mb{r}_t| \mb{x}_t, \mb{H},  \gamma_{d,t})} \right \} \\ \nonumber
    & \propto \mathbbm{1} (\mb{r}_t \in [\mb{y}_t^{\mr{low}}, \mb{y}_t^{\mr{up}}]) \times \exp\left \{ -\lr{\gamma_{d,t}} \| \mb{r}_t - \lr{\mb{H}}\lr{\mb{x}_t}\|^2\right \},
\end{align}
when $T_p + 1 \leq t \leq T$. Here, $q(r_{m,\ell,t})$ is also a complex Gaussian distribution obtained
from bounding $r_{m,\ell,t} \sim \mc{CN}(\lr{[\mb{H}_{\ell}]_{:m}} \lr{\mb{x}_t}, \gamma_{d,t}^{-1})$ to the interval $(y_{m,\ell,t}^{\mr{low}}, y_{m,\ell,t}^{\mr{up}})$. Later, in the Appendix, we will use $\ms{U}$ and $\ms{V}$ functions to represent the variational mean $\lr{r_{m,\ell,t}}$ and variance
$\tau^{r}_{m,\ell,t}$, respectively.

%------- Q-E Algorithm ------- %
\begin{algorithm}[t]
    \small
    \textbf{Input:} $\mb{Y}_p$, $\mb{Y}_d$, $\mb{X}_p$, $\bs{\Sigma}$, $I_{\mr{tr}}$, prior distribution of $p_a, \forall a \in \mc{S}$, $a_p, b_p$, and $a_{d,t}, b_{d,t}, t\in[T_p+1,T]$ \\
    \textbf{Output:} $\hat{\mb{H}}$, $\hat{\mb{X}}_d$ \\
    \SetAlgoNoLine
    Set $\mb{R}_p = \mb{Y}_p, \mb{R}_d = \mb{Y}_d, \hat{\mb{H}} = \mb{0}, \hat{\mb{X}}_d = \mb{0}$. \\
    Initialize $\tau^{r}_{m,\ell,t} = 0, ~ \forall m \forall \ell \forall t$. \\
        \Repeat{convergence}{
        Obtain $\lr{\gamma_p}$ via \eqref{eq:q(gamma_p)}.\\
        \For{$t=1,2,\ldots,T_p$}{
        Find $\lr{r_{m,\ell,t}}, \forall m \forall \ell$, using \eqref{eq:q_r_p} and \eqref{eq:F_appndx}.\\
        Get $\tau^{r}_{m,\ell,t}, \forall m \forall \ell$, using \eqref{eq:q_r_p} and \eqref{eq:G_appndx}.\\
        }
        \For{$t=T_p+1,\ldots,T$}{
        Calculate $\lr{\gamma_{d,t}}$ using \eqref{eq:q(gamma_d)}.\\
        Attain $\lr{r_{m,\ell,t}}, \forall m \forall \ell$, using \eqref{eq:q_r_d} and \eqref{eq:F_appndx}.\\
        Obtain $\tau^{r}_{m,\ell,t}, \forall m \forall \ell$, using \eqref{eq:q_r_d} and \eqref{eq:G_appndx}.
        }
        Apply \eqref{mean-h} and \eqref{Sigma-h} to get $\blr{{\mb{h}}_{i,\ell}}$ and $\hat{\bs{\Sigma}}_{i,\ell}, \forall i \forall \ell$, respectively. \\
            \For{$t=T_p+1,\ldots,T$}{
            \For{$i=1,2,\ldots,K$}{
            Get $q_i(x_{i,t})$ as in \eqref{x-t-i}.\\
                    Calculate $\lr{x_{i,t}}$ and $\tau^x_{i,t}$ based on~\eqref{eq:mean_x_i_t} and~\eqref{eq:var_x_i_t}.\\
            }
            }
        Set $r_{m,\ell,t} = \lr{r_{m,\ell,t}}, \mb{h}_{i,\ell} = \lr{\mb{h}_{i,\ell}}, \bs{\Sigma}_{i,\ell} = \hat{\bs{\Sigma}}_{i,\ell}$. \\
        Set $x_{i,t} = \lr{x_{i,t}}, \gamma_p = \lr{\gamma_p}, \gamma_{d,t} = \lr{\gamma_{d,t}}$.
    }
            
Compute $\hat{x}_{i,t} = \argmax_{a\in \mc{S}} q_i(a), \forall i~\text{and}~ t\in[T_p+1, T]$.\\
Calculate $\hat{\mb{H}} = \mb{H}$.
    \caption{VB(Q-E) Method}
    \label{algo-1}
\end{algorithm}

Algorithm~\ref{algo-1} presents a pseudocode for performing the JED process using our proposed VB method in the Q-E scenario, where $I_{\mr{tr}}$ denotes the number of iterations in the CAVI algorithm. We will discuss the fronthaul signaling overhead required in the Q-E scenario in detail later in Section \ref{subsect:fronthaul_signaling}.

It is important to mention that the PFL scenario follows the steps outlined in Algorithm~\ref{algo-1}, except for steps $7$ to $9$, $12$, and $13$, which are skipped.
% ---------------
\subsection{Scenario 3 -- Estimation-and-Quantization}

In the two previous scenarios, the CPU is responsible for the entire process. However, it could be possible to leverage APs to partially process the received signals, then forward the processed signals to the CPU, which would complete the remaining processing. To achieve this, we introduce the E-Q scenario, which includes two parts: \emph{i.)} AP pre-processing in the pilot transmission phase, and \emph{ii.)} CPU processing during the data transmission phase. In particular, the $\ell^{\mr{th}}$ AP performs local processing relying only on its received signal, $\mb{r}_{\ell,t}, 1 \leq t \leq T_p,$  to find $\mb{h}^{[\mr{loc}]}_{i,\ell}$ as a local estimate of $\mb{h}_{i,\ell}$. Then, it computes $\mb{h}^{\mr{q}}_{i,\ell} = \mc{Q}(\mb{h}^{\mr{loc}}_{i,\ell})$ as a quantized version of $\mb{h}^{\mr{loc}}_{i,\ell}$ and forwards it to the CPU.  Next, in the data transmission phase, the $\ell^{\mr{th}}$ AP sends $\mb{h}^{\mr{q}}_{i,\ell}$ and $\mb{y}_{\ell,t}$ to the CPU and lets the CPU follow the VB framework to do JED based on $\mb{H}^{\mr{q}}$ and $\mb{Y}_d$, where $\mb{H}^{\mr{q}}$ is the stacked matrix containing all $\mb{h}^{\mr{q}}_{i,\ell},  \forall i, \forall \ell $. In the following, we explain the details of the VB approach for AP pre-processing and CPU processing tasks.

{\bf AP pre-processing:} In this part, the $\ell ^{\mathrm{th}}$ AP performs CE using its unquantized received signal $\mb{r}_{\ell,t}$. To do so, it follows the CAVI algorithm to update unknown variables $\mb{h}^{\mr{loc}}_{i,\ell}$ and $\gamma^{\mr{loc}}_{p,\ell}$, where $\gamma^{\mr{loc}}_{p,\ell}$ is the local estimate of $\gamma_p$ at the $\ell ^{\mr{th}}$ AP.

\subsubsection{ Updating $\mb{h}^{\mr{loc}}_{i,\ell}$} By computing the expectation of~\eqref{conditional} for all latent variables except for $\mb{h}_{i,\ell}$, and by taking the point into account that only $\mb{r}_{\ell,t}$ is available the $\ell^{\mr{th}}$ AP, we express the variational distribution $q(\mb{h}^{\mr{loc}}_{i,\ell})$ as~\eqref{q-h-i_E&Q} (at the top of the next page), 
\begin{figure*}
    \begin{align} 
    \label{q-h-i_E&Q}
        q(\mb{h}^{\mr{loc}}_{i,\ell}) & \propto \mr{exp}\Big\{\big\langle \ln \prod_{t=1}^{T_p}p(\mb{r}_{\ell,t}|\mb{H}_{\ell}, \gamma^{\mr{loc}}_{p,\ell}; \mb{x}_t) + \ln p(\mb{h}_{i,\ell};\bs{\Sigma}_{i,\ell}) \big\rangle\Big\}  \\
        & \propto \mr{exp}\Big\{- \blr{\gamma^{\mr{loc}}_{p,\ell} \sum_{t=1}^{T_p}\big\|\mb{r}_{\ell,t}-\mb{H}_{\ell}\mb{x}_t\big\|^2} -  \blr{\mb{h}_{i,\ell}^H \bs{\Sigma}^{-1}_{i,\ell} \mb{h}_{i,\ell}}  \Big\} \nonumber \\
        & \propto \mr{exp}\bigg\{-\mb{h}_{i,\ell}^H\Big[\lr{\gamma^{\mr{loc}}_{p,\ell}}\Big[\sum_{t=1}^{T_p}|x_{i,t}|^2\Big]\mb{I}_M+ \bs{\Sigma}^{-1}_{i,\ell}\Big]\mb{h}_{i,\ell}  + 2\,\Re\Big\{\lr{\gamma^{\mr{loc}}_{p,\ell}}\mb{h}_{i,\ell}^H \Big[\sum_{t=1}^{T_p}\big(\mb{r}_{\ell,t} - \sum_{j=1, j\neq i}^K \lr{\mb{h}_{j,\ell}} x_{j,t}\big)x_{i,t}^*\Big]\Big\}\bigg\}. \nonumber 
    \end{align}
    \hrulefill
    \vspace{-4mm}
\end{figure*}
which is Gaussian with the subsequent covariance matrix and mean: 
\begin{align}
\label{Sigma-h_E&Q}
    \hat{\bs{\Sigma}}^{\mr{loc}}_{i,\ell}  &=\Big[\lr{\gamma^{\mr{loc}}_{p,\ell}}\Big[\sum_{t=1}^{T_p}|x_{i,t}|^2\Big]\mb{I}_M + \bs{\Sigma}^{-1}_{i,\ell}\Big]^{-1},\\
\label{mean-h_E&Q}
    \lr{\mb{h}^{\mr{loc}}_{i,\ell}} &=\hat{\bs{\Sigma}}^{\mr{loc}}_{i,\ell}\Bigg[\lr{\gamma^{\mr{loc}}_{p,\ell}}\Big[ \sum_{t=1}^{T_p}\big(\mb{r}_{\ell,t} - \sum_{j=1, j\neq i}^K \lr{\mb{h}_{j,\ell}} x_{j,t}\big)x_{i,t}^* \Big] \Bigg].
\end{align}

\subsubsection{Updating $\gamma^{\mr{loc}}_{p,\ell}$} We take the expectation of~\eqref{conditional} with respect to all latent variables except for $\gamma_p$, to derive the variational distribution $q(\gamma^{\mr{loc}}_{p,\ell})$ as follows:
\begin{align} 
\label{eq:q(gamma_p)_E&Q_ini}
    q(\gamma^{\mr{loc}}_{p,\ell}) & \propto \mr{exp}\Big\{\big\langle \sum_{t=1}^{T_p}\ln p(\mb{r}_{\ell,t}|\mb{H}_{\ell},\gamma^{\mr{loc}}_{p,\ell}; \mb{x}_t)  + \ln p(\gamma^{\mr{loc}}_{p,\ell})\big\rangle\Big\}  \nonumber \\
    &\propto \mr{exp} \Big\{M T_p\ln \gamma^{\mr{loc}}_{p,\ell} - \gamma^{\mr{loc}}_{p,\ell} \blr{\sum_{t=1}^{T_p}\|\mb{r}_{\ell,t}-\mb{H}_{\ell}\mb{x}_t\|^2} \nonumber \\ & + (a_p-1)\ln \gamma^{\mr{loc}}_{p,\ell} - b_p \gamma^{\mr{loc}}_{p,\ell} \Big\},
\end{align}
which leads to 
\begin{align} 
\label{eq:q(gamma_p)_E&Q}
\lr{\gamma^{\mr{loc}}_{p,\ell}} = \frac{a_p + MT_p}{b_p + \sum_{t=1}^{T_p}\blr{\|\mb{r}_{\ell,t}-\mb{H}_{\ell}\mb{x}_t\|^2}},
\end{align} where $
\blr{\|\mb{r}_{\ell,t}-\mb{H}_{\ell}\mb{x}_t\|^2} = \|\mb{r}_{\ell,t}-\lr{\mb{H}^{\mr{loc}}_{\ell}}\mb{x}_t\|^2 + \sum_{i=1}^K|x_{i,t}|^2\tr\{\hat{\bs{\Sigma}}^{\mr{loc}}_{i,\ell}\}$ using Lemma \ref{theorem-2}, and $\mb{H}^{\mr{loc}}_{\ell}=\left[\mb{h}^{\mr{loc}}_{1,\ell},\mb{h}^{\mr{loc}}_{2,\ell},\ldots,\mb{h}^{\mr{loc}}_{K,\ell}\right]$.

{\bf CPU processing:} In this part, the CPU utilizes the information sent by all APs (i.e., $\mb{H}^{\mr{q}}$ and $\mb{Y}_d$) and applies the VB method to obtain the variational distribution $q(\mb{R}_d,\mb{X}_d,\mb{H},\bs{\gamma}_d)$ to approximate the posteriori distribution $p(\mb{R}_d,\mb{X}_d,\mb{H}, \bs{\gamma}_d| \mb{H}^{\mr{q}}, \mb{Y}_d; \bs{\Sigma})$ as below:
\begin{align}
    &p(\mb{R}_d,\mb{X}_d,\mb{H},\bs{\gamma}_{d}| \mb{H}^{\mr{q}}, \mb{Y}_d; \bs{\Sigma})\approx q(\mb{R}_d,\mb{X}_d,\mb{H},\bs{\gamma}_{d}) \\  \nonumber 
    &\hspace{-1mm}= \left[\prod_{t=T_p+1}^{T} \hspace{-3mm}q(\mb{r}_t)\right] \hspace{-1mm}  \left[\prod_{i=1}^K \prod_{\ell=1}^L 
    q(\mb{h}_{i,\ell})\right]\hspace{-1mm} \left[ \prod_{t=T_p + 1}^{T} \prod_{i=1}^K q_i(x_{i,t})q(\gamma_{d,t})\right].
\end{align}

Next, we leverage the CAVI algorithm to find the local optimal solutions for the variational distributions $q(\mb{r}_t)$, $q(\mb{h}_{i,\ell})$, and $q(x_{i,t})$, and $q(\gamma_{d,t})$. To achieve this, we need to compute the joint probability, which is given by \eqref{eq:conditional_prob_E&Q} on the next page, where $\mb{H}^{\mr{loc}}$ is the stacked matrix containing all $\mb{h}^{\mr{loc}}_{i,\ell}, \forall i \forall \ell$.
\begin{figure*}
    \begin{align}
\label{eq:conditional_prob_E&Q}
        & p(\mb{Y}_d,\mb{R}_d, \mb{X}_d, \mb{H}, \mb{H}^{\mr{q}}, \mb{H}^{\mr{loc}}, \bs{\gamma}_d; \bs{\Sigma}) = p(\mb{Y}_d|\mb{R}_d) p(\mb{R}_d|\mb{X}_d, \mb{H}, \bs{\gamma}_d) p(\mb{H}^{\mr{q}}| \mb{H}^{\mr{loc}}) p(\mb{H}^{\mr{loc}} | \mb{H}) p(\mb{H}; \bs{\Sigma}) p(\mb{X}_d) p(\bs{\gamma}_d)\\  
        & \quad \quad = \left[\prod_{t=T_p + 1}^{T}p(\mb{y}_t|\mb{r}_t) p(\mb{r}_t|\mb{x}_t, \mb{H}, \gamma_{d,t}) p(\mb{x}_t)p(\gamma_{d,t})\right] p(\mb{H}^{\mr{q}}| \mb{H}^{\mr{loc}}) p(\mb{H}^{\mr{loc}} | \mb{H}) p(\mb{H}; \bs{\Sigma})  \nonumber \\
        & \quad \quad =\left[\prod_{t=T_p + 1}^{T}  \left( \prod_{m=1}^M\prod_{\ell=1}^L p(y_{m,\ell,t}|r_{m,\ell,t})p(r_{m,\ell,t}|\mb{x}_t, [\mb{H}_\ell]_{:m}, \gamma_{d,t})\right) \left(\prod_{t=T_p+1}^{T}\prod_{i=1}^{K} p(x_{i,t})\right) \right]\left[\prod_{t=T_p+1}^{T}p(\gamma_{d,t})\right] \nonumber \\
        & \quad \quad \times \left[\prod_{i=1}^{K}  \prod_{\ell=1}^{L}  p(\mb{h}^{\mr{q}}_{i,\ell}|\mb{h}^{\mr{loc}}_{i,\ell}) p(\mb{h}^{\mr{loc}}_{i,\ell}|\mb{h}_{i,\ell})p(\mb{h}_{i,\ell}; \bs{\Sigma}_{i,\ell})\right] 
        \nonumber.
    \end{align}
    \hrulefill
    \vspace{-5mm}
\end{figure*}

In this part, the updating procedures are analogous to what we presented in Section \ref{subsect:Q&E}, except for $\mb{h}_{i,\ell}$, by ignoring the information corresponding to the pilot transmission phase. Thus, we only provide the details about finding $q(\mb{h}_{i,\ell})$.

\subsubsection{Updating $\mb{h}_{i,\ell}$}
We compute the expectation of~\eqref{eq:conditional_prob_E&Q} for all latent variables except for $\mb{h}_{i,\ell}$. We express the variational distribution $q(\mb{h}_{i,\ell})$ as~\eqref{q-h-i_CPU_E&Q} (on the next page), 
\begin{figure*}
    \begin{align} 
    \label{q-h-i_CPU_E&Q}
        q(\mb{h}_{i,\ell}) &\propto \mr{exp}\Big\{\big\langle  \ln \prod_{T_p + 1}^{T}p(\mb{r}_t|\mb{x}_t, \mb{H},\gamma_{d,t}) + \ln p(\mb{h}^{\mr{q}}_{i,\ell}|\mb{h}^{\mr{loc}}_{i,\ell}) + \ln p(\mb{h}^{\mr{loc}}_{i,\ell}|\mb{h}_{i,\ell}) + \ln p(\mb{h}_{i,\ell};\bs{\Sigma}_{i,\ell}) \big\rangle\Big\}  \nonumber \\
        &\propto p(\mb{h}^{\mr{q}}_{i,\ell}|\mb{h}^{\mr{loc}}_{i,\ell})  p(\mb{h}^{\mr{loc}}_{i,\ell}|\mb{h}_{i,\ell}) \times \mr{exp}\Big\{- \blr{ \sum_{t=T_p + 1}^{T}\gamma_{d,t} \big\|\mb{r}_t-\mb{H}\mb{x}_t\big\|^2} -  \blr{ \mb{h}_{i,\ell}^H\bs{\Sigma}^{-1}_{i,\ell}\mb{h}_{i,\ell}}  \Big\} \nonumber \\
        &\propto p(\mb{h}^{\mr{q}}_{i,\ell}|\mb{h}^{\mr{loc}}_{i,\ell}) p(\mb{h}^{\mr{loc}}_{i,\ell}|\mb{h}_{i,\ell}) \times  \mc{CN}(\mb{h}_{i,\ell}; \tilde{\mb{h}}_{i,\ell}, \tilde{\bs{\Sigma}}_{i,\ell}).
    \end{align}
    \hrulefill
    \vspace{-5mm}
\end{figure*}
where $\tilde{\mb{h}}_{i,\ell}$ and $ \tilde{\bs{\Sigma}}_{i,\ell}$ are given by:
\begin{align}
\label{Sigma-h_CPU_E&Q}
    \tilde{\bs{\Sigma}}_{i,\ell} &= \Big[ \big(\sum_{t= T_p + 1}^{T}\lr{\gamma_{d,t}}\lr{|x_{i,t}|^2} \big) \mb{I}_{M} + \bs{\Sigma}^{-1}_{i,\ell}\Big]^{-1}, \\
\label{mean-h_CPU_E&Q}
    \tilde{\mb{h}}_{i,\ell} &=\tilde{\bs{\Sigma}}_{i,\ell}\Big[  \sum_{t=T_p + 1}^{T}\lr{\gamma_{d,t}}\Big(\mb{r}_{\ell,t} - \sum_{j=1, j\neq i}^K \lr{\mb{h}_{j,\ell}} \lr{x_{j,t}} \Big)\lr{x_{i,t}^*} \nonumber \\
    & \hspace{-5mm}+ \sum_{t= T_p + 1}^{T}\lr{\gamma_{d,t}}\Big(\sum_{\ell^{\prime} = 1, \ell^{\prime} \neq \ell}^{L} \left(\mb{r}_{\ell^{\prime},t} - \lr{\mb{H}_{\ell^{\prime}}} \lr{\mb{x}_t}\right)\Big)\lr{x_{i,t}^*} \Big].
\end{align}

It is essential to note that since $\mb{h}^{\mr{loc}}_{i,\ell}$ is estimated from local observations at the $\ell^{\mr{th}}$ AP, it typically differs from the true channel $\mb{h}_{i,\ell}$. We model this mismatch as $\mb{h}^{\mr{loc}}_{i,\ell} = \mb{h}_{i,\ell} + \mb{e}_{\mr{h},i,\ell}$, where $\mb{e}_{\mr{h},i,\ell}$ is the corresponding CE error, and each element of $\mb{e}_{\mr{h},i,\ell}$ is 
assumed to be i.i.d. zero-mean complex Gaussian with known variance $N^{\mr{e}}_{i,\ell}$. Under this model, \eqref{q-h-i_CPU_E&Q} describes the distribution of a quantized version of a noisy Gaussian variable. Moreover, the variational distribution $q(\mb{h}_{i,\ell})$ is inherently separable as $\prod_{m=1}^M \prod_{i=1}^K\prod_{\ell=1}^L q(h_{m,i,\ell})$. Therefore, the corresponding mean and variance of $\Re\{h_{m,i,\ell}\}$, denoted by $\lr{\Re\{h_{m,i,\ell}\}}$ and $[\hat{\bs{\Sigma}}^{[1]}_{i,\ell}]_{mm}$, respectively, can be computed using the approach proposed in \cite{wen2015bayes} as \eqref{eq:mean_real_h} and \eqref{eq:var_real_h}, shown on the next page,
\begin{figure*}
\begin{align}
\label{eq:mean_real_h}
\lr{\Re\{h_{m,i,\ell}\}} &= \Re\{\tilde{h}_{m,i,\ell}\} + \frac{\mr{sign}\big(\Re\{h^{\mr{q}}_{m,i,\ell}\}\big)[\tilde{\bs{\Sigma}}_{i,\ell}]_{mm}}{\sqrt{2(N^{\mr{e}}_{i,\ell} + [\tilde{\bs{\Sigma}}_{i,\ell}]_{mm})}} \left( \frac{u(\eta_1) - u(\eta_2)}{U(\eta_1) - U(\eta_2)} \right),\\
\label{eq:var_real_h}
[\hat{\bs{\Sigma}}^{[1]}_{i,\ell}]_{mm} &= \frac{[\tilde{\bs{\Sigma}}_{i,\ell}]_{mm}}{2} - \frac{\left([\tilde{\bs{\Sigma}}_{i,\ell}]_{mm}\right)^2}{2(N^{\mr{e}}_{i,\ell} + [\tilde{\bs{\Sigma}}_{i,\ell}]_{mm})} \left( \frac{\eta_1 u(\eta_1) - \eta_2 u(\eta_2)}{U(\eta_1) - U(\eta_2)} + \left( \frac{u(\eta_1) - u(\eta_2)}{U(\eta_1) - U(\eta_2)} \right)^2 \right),\\
\label{eq:eta1_real_h}
\eta_1 & = \frac{\mr{sign}\big(\Re\{h^{\mr{q}}_{m,i,\ell}\}\big)\Re\{\tilde{h}_{m,i,\ell}\} - \min\{ |\Re \{h_{m,\ell}^{\mr{q}, \mr{low}}\}|, |\Re\{h_{m,\ell}^{\mr{q}, \mr{up}}\}| \}}{\sqrt{(N^{\mr{e}}_{i,\ell} + [\tilde{\bs{\Sigma}}_{i,\ell}]_{mm})/2}},  \\
\label{eq:eta2_real_h}
\eta_2 &= \frac{\mr{sign}\big(\Re\{h^{\mr{q}}_{m,i,\ell}\}\big)\Re\{\tilde{h}_{m,i,\ell}\} - \max\{ |\Re \{h_{m,\ell}^{\mr{q}, \mr{low}}\}|, |\Re\{h_{m,\ell}^{\mr{q}, \mr{up}}\}| \}}{\sqrt{(N^{\mr{e}}_{i,\ell} + [\tilde{\bs{\Sigma}}_{i,\ell}]_{mm})/2}}, 
\end{align}
    \hrulefill
    \vspace{-5mm}
\end{figure*} where $u(x)$ and $U(x)$ represent the PDF and CDF of the random variable $x$, respectively.

Similar to \eqref{eq:mean_real_h}, \eqref{eq:var_real_h}, \eqref{eq:eta1_real_h}, and \eqref{eq:eta2_real_h}, by replacing the real parts with imaginary parts, we can derive $\lr{\Im\{h_{m,i,\ell}\}}$ and $[\hat{\bs{\Sigma}}^{[2]}_{i,\ell}]_{mm}$ as the corresponding mean and variance of $\Im\{h_{m,i,\ell}\}$, respectively. Finally, $\lr{h_{m,i,\ell}} = \lr{\Re\{h_{m,i,\ell}\}} + j\lr{\Im\{h_{m,i,\ell}\}}$ and $[\hat{\bs{\Sigma}}_{i,\ell}]_{mm}= [\hat{\bs{\Sigma}}^{[1]}_{i,\ell}]_{mm} + [\hat{\bs{\Sigma}}^{[2]}_{i,\ell}]_{mm}$.

% It is essential to note that the variational distribution $q(\mb{h}_{i,\ell})$ is inherently separable as $\prod_{m=1}^M \prod_{i=1}^K\prod_{\ell=1}^L q(h_{m,i,\ell})$ and the variational distribution
% $q(h_{m,i,\ell})$ is the complex Gaussian distribution obtained
% from bounding $h_{m,i,\ell} \sim \mc{CN}([\tilde{\mb{h}}_{i,\ell}]_m, [\tilde{\bs{\Sigma}}_{i,\ell}]_{m,m})$ to the interval $(h_{m,\ell}^{\mr{q}, \mr{low}}, h_{m,\ell}^{\mr{q}, \mr{up}})$. We follow $\ms{U}$ and $\ms{V}$ functions, described in the Appendix to compute the mean and variance of $h_{m,i,\ell}$, respectively.

Algorithm~\ref{algo-2} provides a pseudocode for implementing the VB(E-Q) method to carry out the JED process. The AP pre-processing is detailed in steps $3$ through $10$, while the CPU processing is performed in steps $11$ through $28$. In the next section, we will provide a detailed discussion on the fronthaul signaling overhead requirements for the E-Q scenario.
%------- E-Q Algorithm ------- %
\begin{algorithm}[t]
    \small
    \textbf{Input:} $\mb{Y}_p$, $\mb{Y}_d$, $\mb{X}_p$, $\bs{\Sigma}$, $I_{\mr{tr}}$, $N^{\mr{e}}_{i,\ell}, \forall i \forall \ell$, prior distribution of $p_a, \forall a \in \mc{S}$, $a_p, b_p$, and $a_{d,t}, b_{d,t}, t\in[T_p+1,T]$\\
    \textbf{Output:} $\hat{\mb{H}}$, $\hat{\mb{X}}_d$ \\
    \SetAlgoNoLine
    %-------- AP Pre-processing ----------%
    \For{$\ell = 1,2, \ldots, L$}{
    Set $ \hat{\mb{H}}^{\mr{loc}}_\ell = \mb{0}$. \\
        \Repeat{convergence}{
        Obtain $\lr{\gamma^{\mr{loc}}_{p,\ell}}$ via \eqref{eq:q(gamma_p)_E&Q}.\\
        Apply \eqref{mean-h_E&Q} and \eqref{Sigma-h_E&Q} to get $\blr{{\mb{h}}^{\mr{loc}}_{i,\ell}}$ and $\hat{\bs{\Sigma}}^{\mr{loc}}_{i,\ell}, \forall i$, respectively. \\
        Set $\mb{h}^{\mr{loc}}_{i,\ell} = \lr{\mb{h}^{\mr{loc}}_{i,\ell}}, \bs{\Sigma}^{\mr{loc}}_{i,\ell} = \hat{\bs{\Sigma}}^{\mr{loc}}_{i,\ell}, \gamma^{\mr{loc}}_{p,\ell} = \lr{\gamma^{\mr{loc}}_{p,\ell}}$.
    }
    Calculate $\hat{\mb{H}}^{\mr{loc}}_\ell = \mb{H}^{\mr{loc}}_\ell$ and $\mb{H}^{\mr{q}}_{\ell} = \mc{Q}(\hat{\mb{H}}^{\mr{loc}}_{\ell})$.\\
    }
    %-------- CPU Processing ----------%
    Set $\mb{R}_d = \mb{Y}_d, \hat{\mb{H}} = \mb{H}^{\mr{q}}, \hat{\mb{X}}_d = \mb{0}$. \\
    Initialize $\tau^{r}_{m,\ell,t} = 0, ~ \forall m \forall \ell \forall t$. \\
        \Repeat{convergence}{
        \For{$t=T_p+1,\ldots,T$}{
        Obtain $\lr{\gamma_{d,t}}$ via  \eqref{eq:q(gamma_d)}.\\
        Attain $\lr{r_{m,\ell,t}}, \forall m \forall \ell$, using \eqref{eq:q_r_d} and \eqref{eq:F_appndx}.\\
        Obtain $\tau^{r}_{m,\ell,t}, \forall m \forall \ell$, using \eqref{eq:q_r_d} and \eqref{eq:G_appndx}.
        }
        Apply \eqref{eq:mean_real_h}  to get $\blr{{\mb{h}}_{i,\ell}}, \forall i \forall \ell$. \\
        Use \eqref{eq:var_real_h} to attain $\hat{\bs{\Sigma}}_{i,\ell}, \forall i \forall \ell$. \\
            \For{$t=T_p+1,\ldots,T$}{
            \For{$i=1,2,\ldots,K$}{
            Get $q_i(x_{i,t})$ as in \eqref{x-t-i}.\\
                    Calculate $\lr{x_{i,t}}$ and $\tau^x_{i,t}$ based on~\eqref{eq:mean_x_i_t} and~\eqref{eq:var_x_i_t}.\\
            }
            }
        Set $r_{m,\ell,t} = \lr{r_{m,\ell,t}}, \mb{h}_{i,\ell} = \lr{\mb{h}_{i,\ell}}, \bs{\Sigma}_{i,\ell} = \hat{\bs{\Sigma}}_{i,\ell}$. \\
        Set $x_{i,t} = \lr{x_{i,t}}, \gamma_{d,t} = \lr{\gamma_{d,t}}$.
    }
            
Compute $\hat{x}_{i,t} = \argmax_{a\in \mc{S}} q_i(a), \forall i~\text{and}~ t\in[T_p+1, T]$.\\
Calculate $\hat{\mb{H}} = \mb{H}$.
    \caption{VB(E-Q) Method}
    \label{algo-2}
\end{algorithm}

        %%%%%%%%%%%%%%%%%%%%%%%%%%%%%%%%%%%%%%%%%%%%%%%%%%%
 	\section{Simulation Results}
        \label{Sec:numerical}
        To assess the performance of our proposed VB methods, in this section, we focus on the uplink scenario in a CF-mMIMO network. We conduct a comparison between our proposed VB methods (i.e., VB(PFL), VB(Q-E), and VB(E-Q)) and both linear and nonlinear JED methods in terms of SER, channel NMSE, computational complexity, and signaling overhead over the fronthaul links.

%For ease of tracking the process, we use Q-E and E-Q to denote the VB(Q-E) and VB(E-Q) methods, respectively, throughout this section.

We evaluate the performance of our VB methods using quadrature phase shift keying (QPSK) modulation. %, assuming the channel covariance matrix of $\bs{\Sigma}_{i,\ell} = \mb{I}_M$. Moreover, 
Unless stated otherwise, we set $L = 8, M= 4, K =16, I_{\mathrm{tr}} = 50$, $T_p = 32$, $T_d = 128$, and $p_a = 1/|\mc{S}|$, where $|\mc{S}|$ denotes the cardinality of $\mc{S}$. Moreover, we normalize the covariance matrix $\bs{\Sigma}_{i,\ell}, \forall i, \forall \ell,$ so that all diagonal elements are equal to $1$. This ensures that $\mathbb{E}[\|\mb{h}_{i,\ell}\|^2] = M$. We then determine the noise variance $N_0$ based on the signal-to-noise ratio (SNR), which is expressed as:
\begin{align}
\mr{SNR} = \frac{\mathbb{E}[\|\mb{H}\mb{x}\|^2]}{\mathbb{E}[\|\mb{n}\|^2]} = \frac{\sum_{i=1}^K \sum_{\ell=1}^L \mr{Tr}(\bs{\Sigma}_{i,\ell})}{MLN_0} = \frac{K}{N_0}.
\end{align}

In this paper, to model the spatial correlation, we adopt the exponential correlation model \cite{loyka2001channel}, applied independently to each column of $\mathbf{H}_{\ell}$. The corresponding covariance matrix $\boldsymbol{\Sigma}_{i,\ell}$ is defined as:
\begin{align}
[\boldsymbol{\Sigma}_{i,\ell}]_{km} = 
\begin{cases} 
 \alpha^{k-m}, & \text{if } k \geq m, \\
 \left( \alpha^{m-k} \right)^*, & \text{if } k < m,
\end{cases}
\end{align}
where $1 \leq k, m \leq M$, and $\alpha$ is the complex correlation coefficient between neighboring antennas. 

We also set $N^{\mr{e}}_{i,\ell}=10^{-2}$ when $\text{SNR}\leq10$ dB, and $N^{\mr{e}}_{i,\ell}=10^{-4}$ for $\text{SNR}>10$ dB. The results are computed by averaging over 100 trials.

\subsection{SER Performance}
In this part, we evaluate the SER performance of our proposed VB methods across various cases, including different SNR levels, number of users, number of APs, pilot transmission length, and data transmission length.

\begin{figure}[t]
\vspace{-6mm}
\centering
  \includegraphics[trim = 0mm 0mm 0mm 0mm, clip, scale=7, width=0.95\linewidth, draft=false]{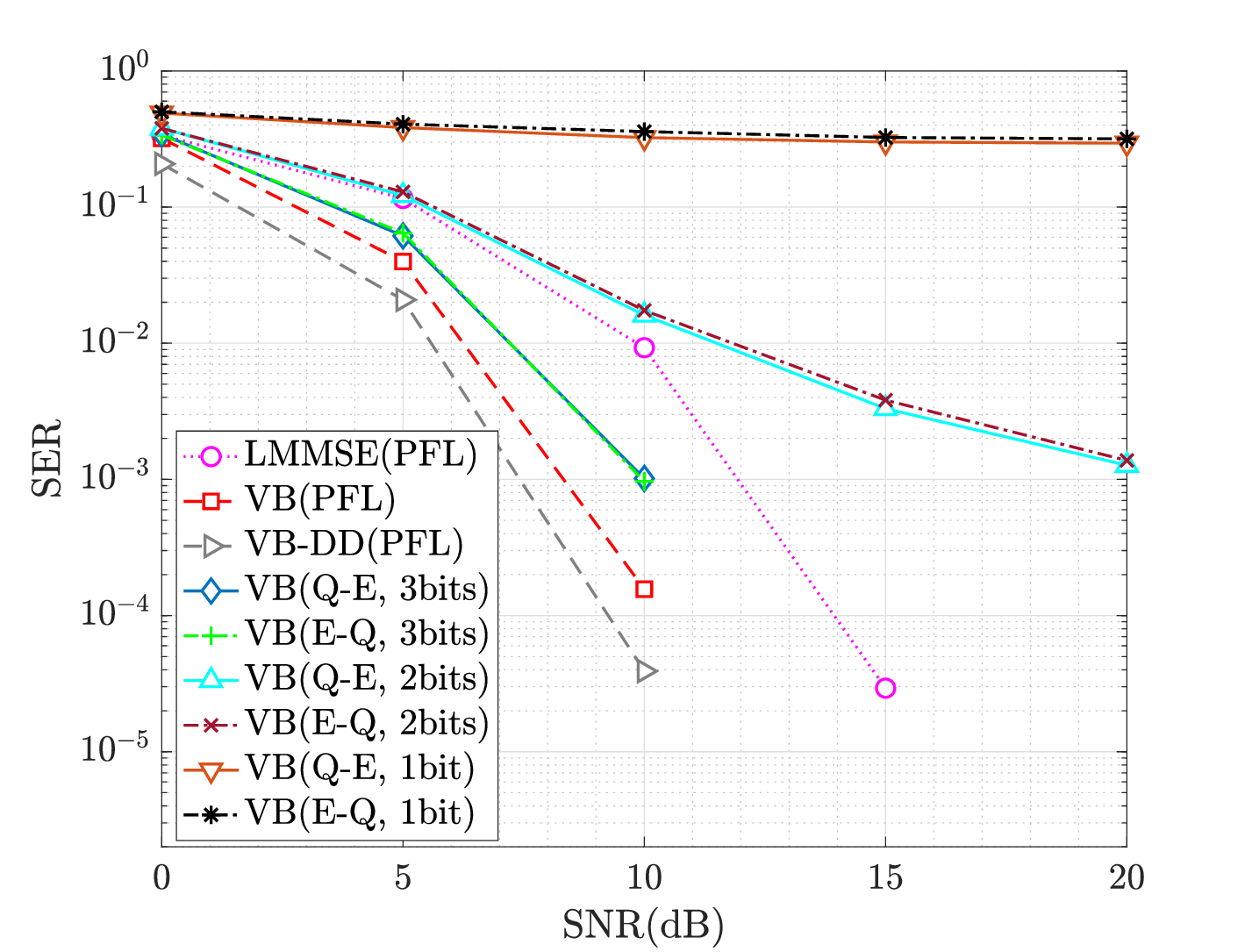}
  \vspace{-3mm}
  \caption{An SER comparison between LMMSE(PFL), VB(PFL), VB-DD(PFL), and different versions of the VB(Q-E) and VB(E-Q) methods under QPSK when $L = 8, M = 4, K = 16, T_p = 32, T_d = 128$, and $\text{SNR}\in [0,20]$ dB.}\label{Fig:SER_vs_SNR}
  \vspace{-5mm}
\end{figure}

Fig. \ref{Fig:SER_vs_SNR} shows the SER as a function of SNR for VB(PFL), LMMSE(PFL), and VB-DD(PFL), as well as VB(Q-E) and VB(E-Q) with i.i.d. channels, evaluated with 1-bit, 2-bit, and 3-bit quantizer configurations, as the SNR ranges from $0$ to $20$ dB. In our VB-based methods, we set $a_p = b_p = a_{d,t} = b_{d,t}=0$. The LMMSE(PFL) approach performs JED by first estimating the channels during the pilot phase using known pilot signals and then utilizing the estimated channels to detect data during the data transmission phase. Furthermore, the VB-DD(PFL) benchmark is a data detection scheme that has access to perfect channel knowledge and PFL, and applies VB inference solely for data detection.

It is evident that the performance of all methods improves with higher SNRs, as the signal becomes stronger in comparison to the noise. Among these, VB-DD(PFL) demonstrates the best performance due to its nonlinear Bayesian model and access to perfect channel knowledge and unquantized received signals from all APs. VB(PFL) ranks second, as it uses unquantized signals but is affected by CE errors inherent to the JED process. The VB(Q-E, 3bits) and VB(E-Q, 3bits) methods take rank third due to their use of the quantized information. %Following these, the VB(E-Q, 3bits) method ranks fourth. Its performance is slightly lower than that of the VB(Q-E) method, consistent with the results reported in \cite{bashar2020uplink}, because of channel estimation errors during the estimation phase, where each AP estimates the channels using only its locally received signals.
Finally, the LMMSE(PFL) method ranks fourth. Despite the CPU having access to received signals from all APs, this method relies on a linear model, resulting in worse performance compared to the VB-based methods with unquantized or 3-bit quantized signals. Notably, VB(PFL), VB(Q-E, 3bits), and VB(E-Q, 3bits) demonstrate gains of about 4 dB, 2 dB, and 2 dB, respectively, at an SER of $10^{-3}$, compared to the LMMSE(PFL) method. 

Additionally, Fig. \ref{Fig:SER_vs_SNR} illustrates that as the resolution of the quantized signals decreases, the performance of both the VB(Q-E) and VB(E-Q) methods deteriorates. For low-resolution cases, the performance curves for these methods saturate at high SNRs. This occurs because, at high SNRs, quantization noise becomes dominant over background noise, rendering further increases in SNR ineffective for performance improvement. Furthermore, as we can see, the performance of VB(E-Q, 2bits) and VB(E-Q, 1bit) is slightly lower than that of their counterparts based on VB(Q-E), consistent with the results reported in \cite{bashar2020uplink}. This is due to CE errors during the estimation phase, where each AP estimates the channels using only its locally received signals.

\begin{figure}[t]
\vspace{-6mm}
\centering
  \includegraphics[trim = 0mm 0mm 0mm 0mm, clip, scale=7, width=0.83\linewidth, draft=false]{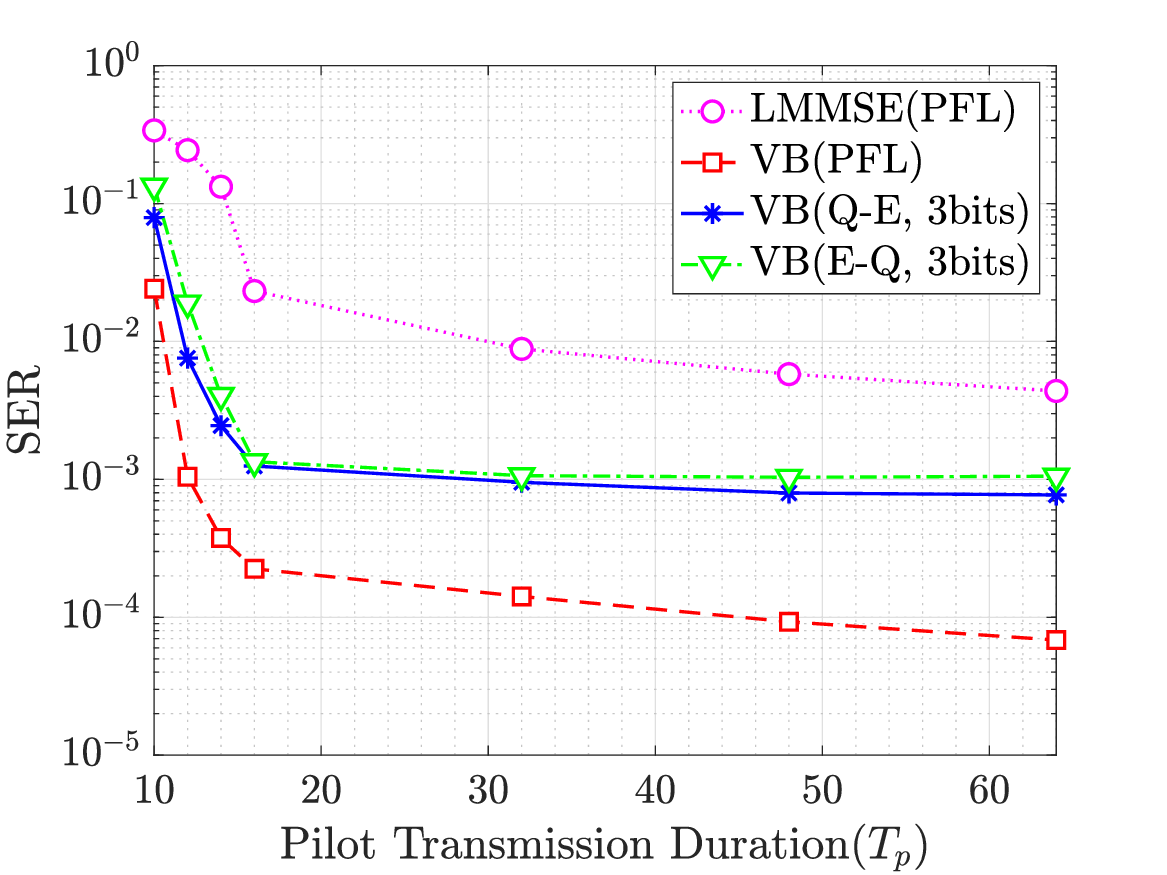}
  \vspace{-3mm}
  \caption{The SER performance comparison between LMMSE(PFL), VB(PFL), VB(Q-E, 3bits), and VB(E-Q, 3bits) utilizing the QPSK modulation, when $L = 8, M = 4, K = 16, T_d=128$, $\text{SNR}=10$ dB, and $T_p\in [10,64]$.}\label{Fig:SER_vs_Tp}
  \vspace{-5mm}
\end{figure}
Fig. \ref{Fig:SER_vs_Tp} presents the SER performance of the LMMSE(PFL), VB(PFL), VB(Q-E, 3bits), and VB(E-Q, 3bits) methods for $T_p \in [10,64]$ with i.i.d channels when $a_p = b_p = a_{d,t} = b_{d,t}=0$. It is worth noting that VB methods do not strictly require orthogonal pilot signals (i.e., $T_p \geq K$). The results show that the performance of all methods improves as $T_p$ increases. Initially, the SER drops rapidly, but the rate of improvement slows down as $T_p$ becomes larger. This happens because, in the JED process during the data transmission phase, the SER performance is influenced by the prior channel knowledge obtained during the pilot phase. When $T_p$ is small, the initial CE is poor due to limited pilot observations, resulting in higher SER. As $T_p$ increases, the prior CE improves, leading to better detection performance. However, once $T_p$ becomes sufficiently large (i.e., $T_p \geq 48$), the performance gain becomes marginal since the JED process iteratively refines the CE during the data transmission phase. Thus, further increasing $T_p$ yields limited additional benefits.

Moreover, VB(Q-E, 3bits) and VB(E-Q, 3bits) reach the saturation point earlier. This is because, after achieving a reasonable CE, their performance becomes limited by quantization noise rather than estimation accuracy. Further, as anticipated, the VB-based methods outperform the LMMSE(PFL) method, benefiting from the nonlinear VB framework.

\begin{figure}[t]
% \vspace{-1mm}
\centering
  \includegraphics[trim = 0mm 0mm 0mm 0mm, clip, scale=7, width=0.83\linewidth, draft=false]{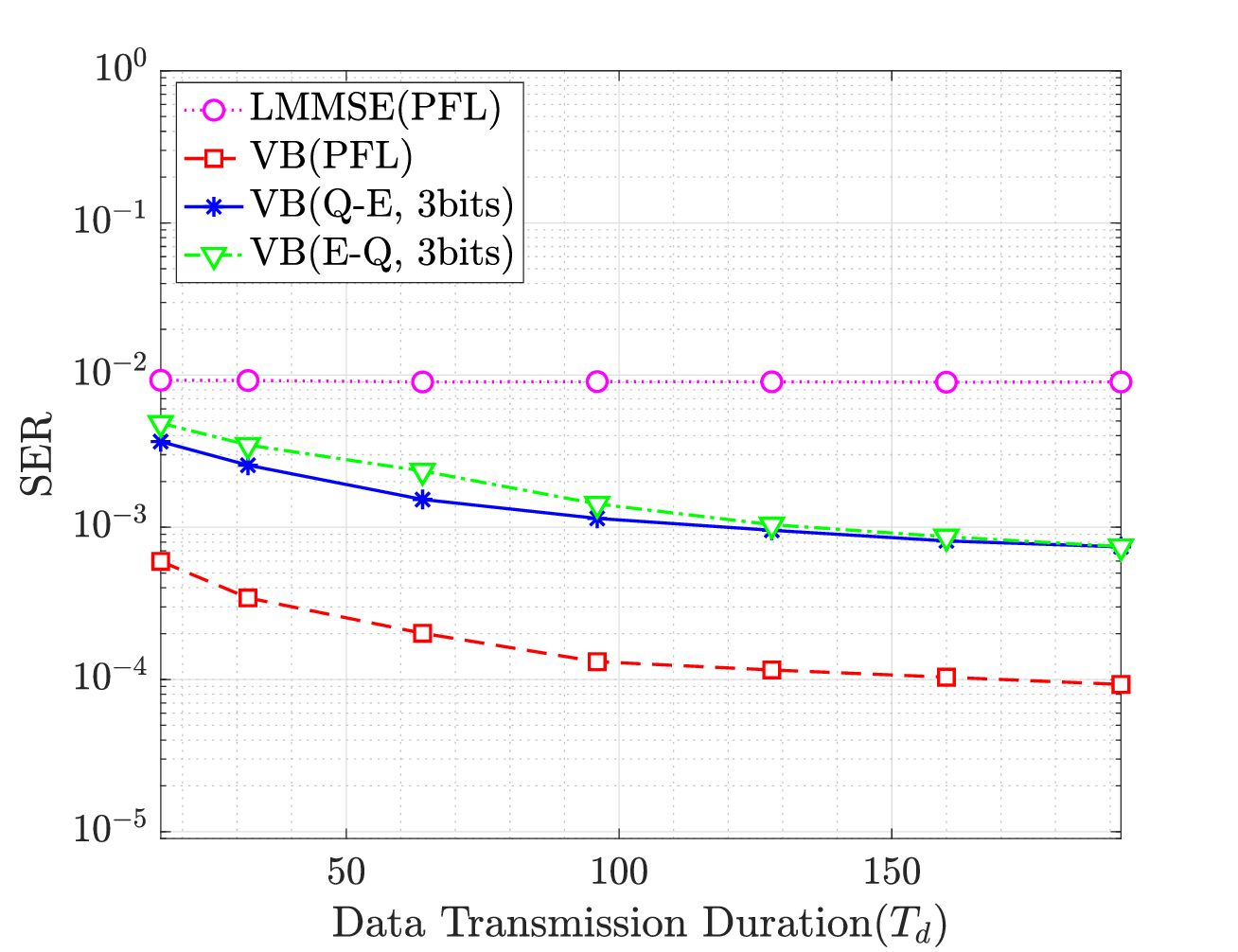}
  \vspace{-3mm}
  \caption{A comparison of SER among LMMSE(PFL), VB(PFL), VB(Q-E, 3bits), and VB(E-Q, 3bits) employing QPSK where $L = 8, M = 4, K = 16, T_p = 32$, $\text{SNR}=10$ dB, and $T_d$ ranging from $16$ to $192$.
  }\label{Fig:SER_vs_Td}
  \vspace{-5mm}
\end{figure}

Next, in Fig. \ref{Fig:SER_vs_Td}, we evaluate the SER performance of LMMSE(PFL), VB(PFL), VB(Q-E, 3bits), and VB(E-Q, 3bits) under i.i.d channels with respect to $T_d$, where $T_d$ varies from $16$ to $192$ and $a_p = b_p = a_{d,t} = b_{d,t}=0$. This figure shows that VB(PFL) outperforms the others due to access to unquantized signals from all APs and the nonlinear VB framework. In contrast, the performance of LMMSE(PFL) remains constant because it only performs CE during the pilot transmission phase and does not involve a recursive optimization process. Consequently, any detection errors that occur during short data transmission periods persist even in scenarios with long $T_d$ values.

\begin{figure}[t]
\vspace{-6mm}
\centering
  \includegraphics[trim = 0mm 0mm 0mm 0mm, clip, scale=7, width=0.83\linewidth, draft=false]{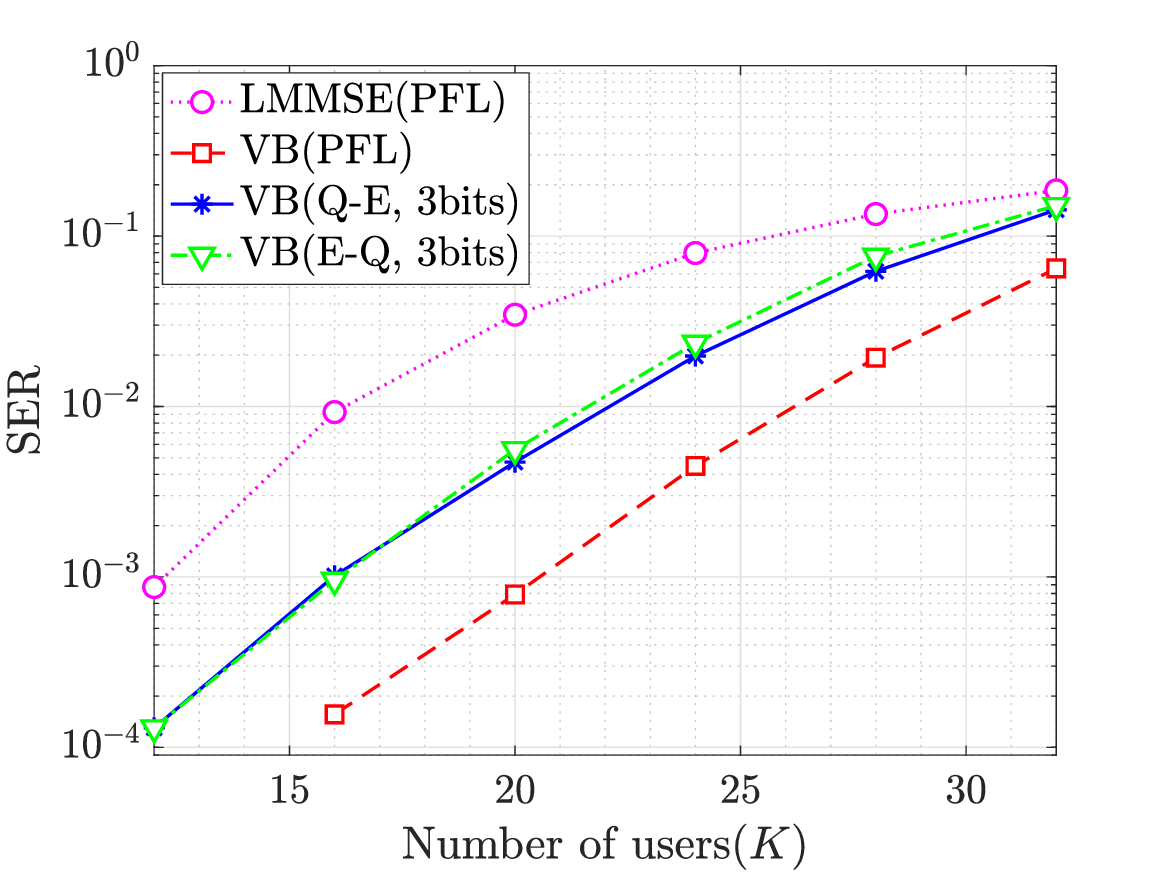}
  \vspace{-3mm}
  \caption{SER performance evaluation of VB(PFL) versus LMMSE(PFL), VB(Q-E, 3bits), and VB(E-Q, 3bits) under the QPSK modulation, when $L = 8, M = 4, T_p = 32, T_d = 128$, $\text{SNR}=10$ dB, and $K \in [12, 32]$.
  }\label{Fig:SER_vs_Users}
  \vspace{-5mm}
\end{figure}

Then, Fig. \ref{Fig:SER_vs_Users} shows the SER as a function of the number of users for LMMSE(PFL) and three variations of our VB-based methods with i.i.d channels, when $K \in [12,32]$ and $a_p = b_p = a_{d,t} = b_{d,t}=0$. As observed, the performance of all methods degrades as $K$ increases. This can be explained to the fact that a network with $K_1$ users can be approximated as one with $K_2$ users, where $K_2 < K_1$, by deactivating $K_1 - K_2$ users. However, when these deactivated users become active, they introduce additional interference to the JED process. The trend in Fig. \ref{Fig:SER_vs_Users} aligns with the previous figures, with VB(PFL) providing the best performance and the VB-based methods consistently outperforming the LMMSE(PFL) method.

\begin{figure}[t]
% \vspace{-2mm}
\centering
  \includegraphics[trim = 0mm 0mm 0mm 0mm, clip, scale=7, width=0.83\linewidth, draft=false]{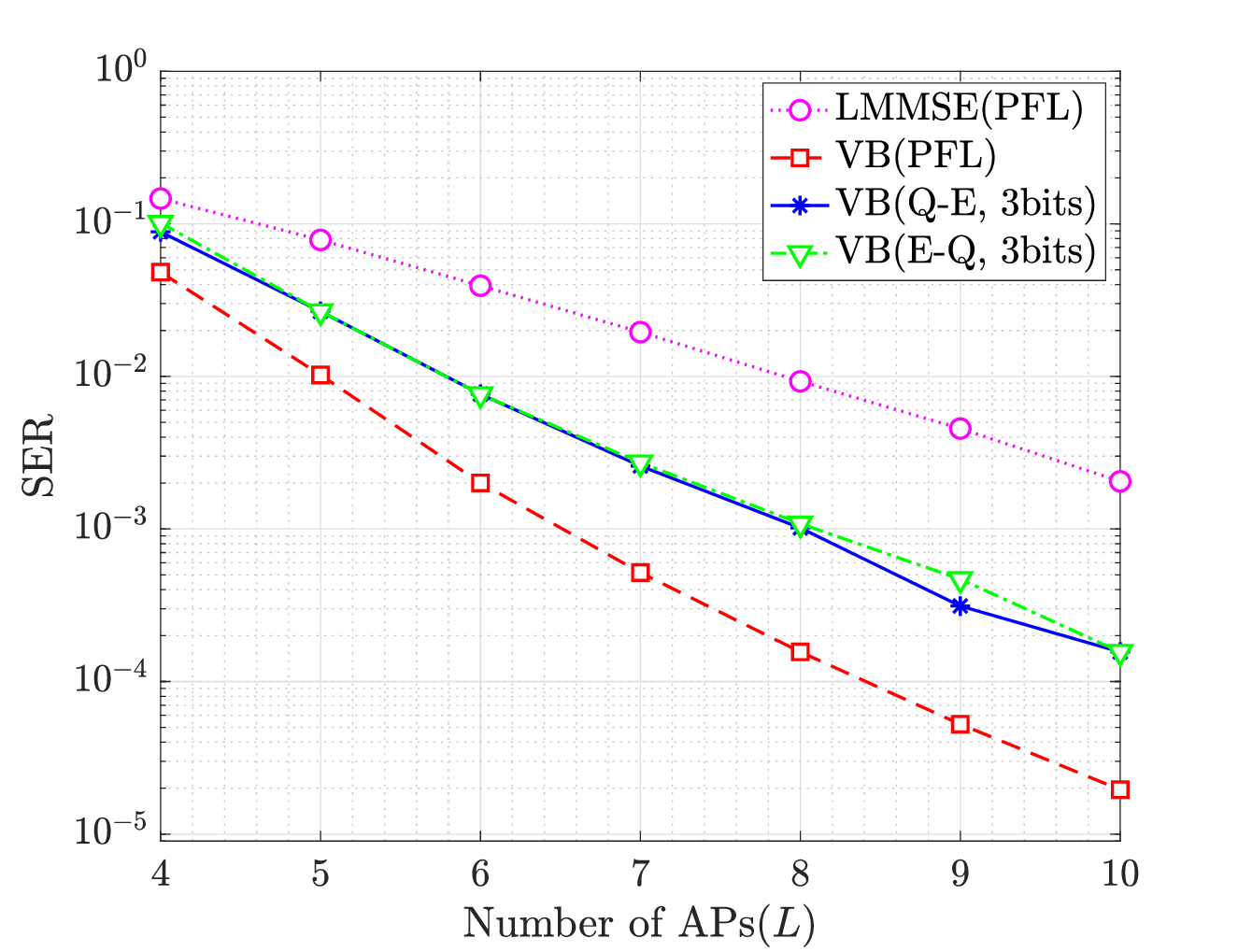}
  \vspace{-3mm}
  \caption{SER comparison of VB-based methods and LMMSE(PFL) in a CF-mMIMO network with QPSK when $M = 4, K =16, T_p = 32, T_d = 128$, $\text{SNR}=10$ dB, and $L$ varies between $4$ and $10$.
  }\label{Fig:SER_vs_APs}
  \vspace{-5mm}
\end{figure}

We also provide an SER analysis of the LMMSE(PFL), VB(PFL), VB(Q-E, 3bits), and VB(E-Q, 3bits) methods with respect to different numbers of APs (i.e., $L \in [4,10]$) in Fig. \ref{Fig:SER_vs_APs} under i.i.d channels with $a_p = b_p = a_{d,t} = b_{d,t}=0$. This figure shows that the SER performance of all methods improves as $L$ increases, due to the availability of a larger number of received signals. Furthermore, our proposed VB-based methods achieve a lower SER compared to LMMSE(PFL), owing to their use of the nonlinear VB framework.

\begin{figure}[t]
\vspace{-6mm}
\centering
  \includegraphics[trim = 0mm 0mm 0mm 0mm, clip, scale=7, width=0.83\linewidth, draft=false]{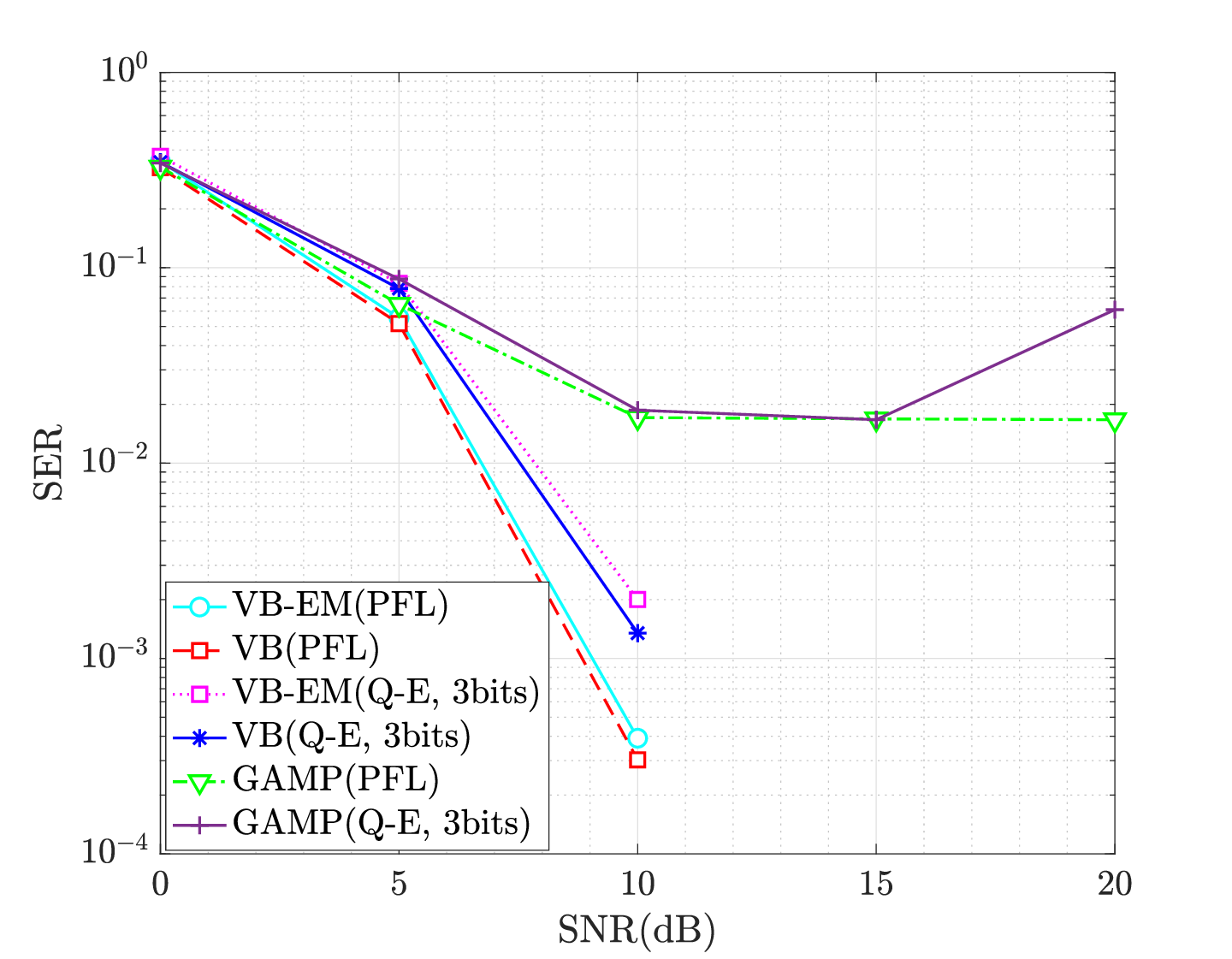}
  \vspace{-3mm}
  \caption{SER comparison among VB-EM(PFL), VB(PFL), VB-EM(Q-E, 3bits), VB(Q-E, 3bits), GAMP(PFL), and GAMP(Q-E, 3bits) under correlated channels using QPSK modulation. The simulation parameters are $L = 8$, $M = 4$, $K = 16$, $T_p = 32$, $T_d = 128$, and $\text{SNR} \in [0, 20]$ dB.}\label{Fig:SER_GAMP_VB_EM}
  \vspace{-5mm}
\end{figure}

Given that our proposed VB-based methods are nonlinear, it is important to evaluate their performance against other nonlinear benchmarks. To this end, we consider the GAMP(PFL) and GAMP(Q-E) methods introduced in \cite{wen2015bayes}, along with the VB-EM(PFL) and VB-EM(Q-E) techniques from \cite{nguyen2024variational}. Note that while these benchmarks were originally developed for mMIMO systems, they can be extended to CF-mMIMO systems. Moreover, since this work is the first to introduce an E-Q approach for the JED problem and no direct counterpart exists in the literature, we limit our comparison to the VB(PFL) and VB(Q-E) methods against the aforementioned nonlinear baselines. Fig.~\ref{Fig:SER_GAMP_VB_EM} presents the SER performance under correlated channels with $\alpha = 0.35 + j0.35$, where the bisection method is used to determine the optimal values of $a_p$, $b_p$, $a_{d,t}$, and $b_{d,t}$. Other settings are configured analogously to Fig.~\ref{Fig:SER_vs_SNR}. Here, VB(PFL) outperforms the other PFL-based methods, while VB(Q-E, 3bits) achieves superior performance than other Q-E-based approaches.
 This improvement is attributed to the fact that GAMP(PFL) and GAMP(Q-E) are not robust under correlated channels and may diverge. Furthermore, while VB-EM(PFL) and VB-EM(Q-E) in \cite{nguyen2024variational} use the EM algorithm to estimate precision, our approach generalizes this by modeling the precision parameters using a Gamma distribution, of which the EM-based solution is a special case.
\subsection{Channel NMSE Performance}
In this part, we focus on the channel NMSE metric to assess the performance of our VB-based methods. We calculate the channel NMSE using the following formula in dB scale.
\begin{align}
    \mr{NMSE~(dB)} = 10 \log_{10} \Big(\frac{\|\mb{H}-\hat{\mb{H}}\|_{\mr{F}}^2}{\|\mb{H}\|_{\mr{F}}^2}\Big).
\end{align}

\begin{figure}[t]
\vspace{-6mm}
\centering
  \includegraphics[trim = 0mm 0mm 0mm 0mm, clip, scale=7, width=0.83\linewidth, draft=false]{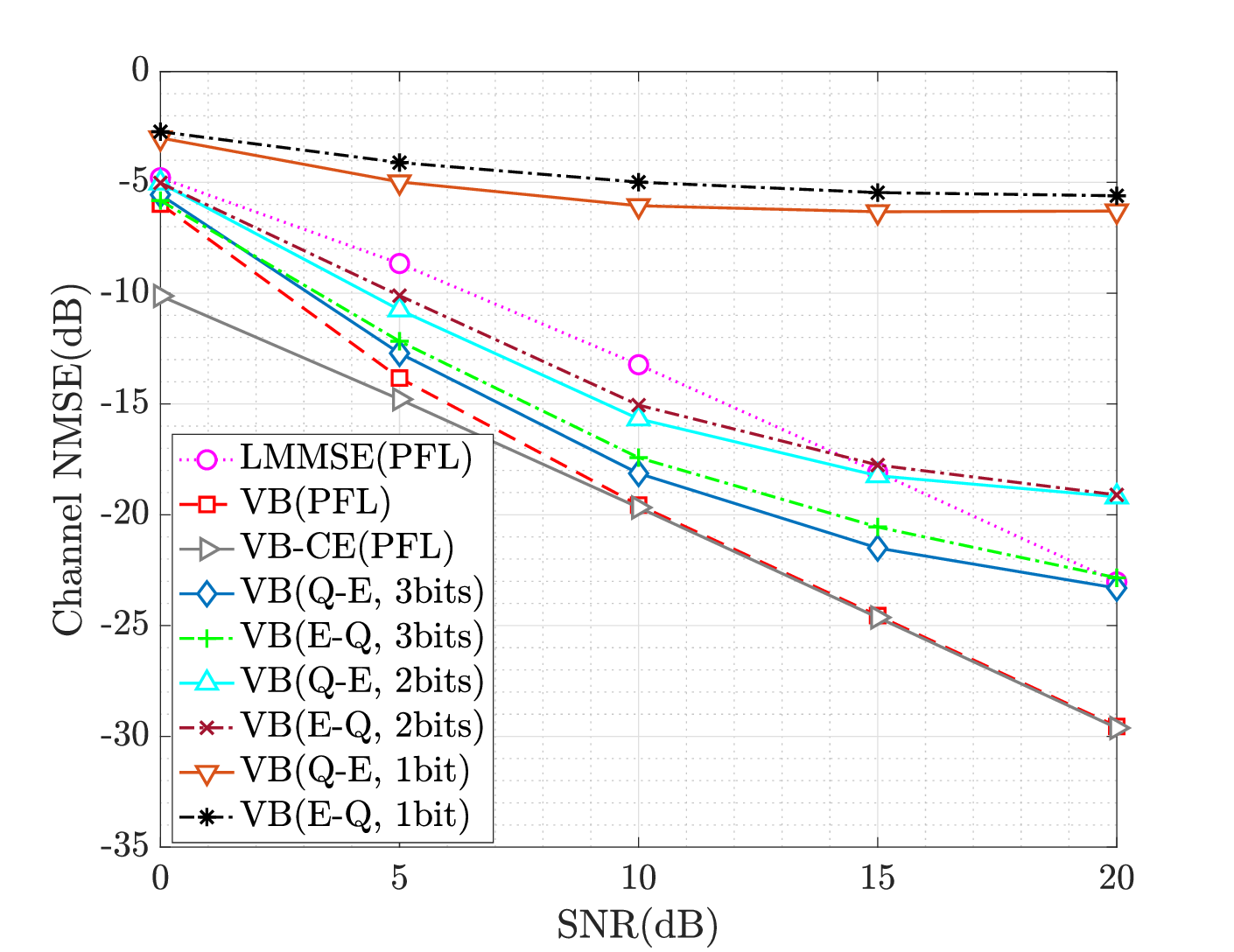}
  \vspace{-3mm}
  \caption{A channel NMSE comparison between LMMSE(PFL), VB(PFL), VB-CE(PFL), and different versions of the VB(Q-E) and VB(E-Q) under QPSK when $L = 8, M = 4, K = 16, T_p = 32, T_d = 128$, and $\text{SNR}\in [0,20]$ dB.}\label{Fig:NMSE_vs_SNR}
  \vspace{-3mm}
\end{figure}

In Fig. \ref{Fig:NMSE_vs_SNR}, we evaluate the channel NMSE performance of our VB-based methods compared to LMMSE(PFL) and VB-CE(PFL), where the simulation settings are the same as in Fig.~\ref{Fig:SER_vs_SNR}. The VB-CE(PFL) method is a CE approach that uses VB inference while treating the entire transmission block as pilot data. Our VB-based methods outperform LMMSE(PFL) for the reasons discussed in Fig. \ref{Fig:SER_vs_SNR}. Further, at low SNRs, a noticeable performance gap exists between our methods and VB-CE(PFL) due to the limited informativeness of data symbols in noisy conditions. However, this gap narrows as the SNR increases, since the data becomes more informative and contributes more effectively to the CE process.

\begin{figure}[t]
% \vspace{-2mm}
\centering
  \includegraphics[trim = 0mm 0mm 0mm 0mm, clip, scale=7, width=0.83\linewidth, draft=false]{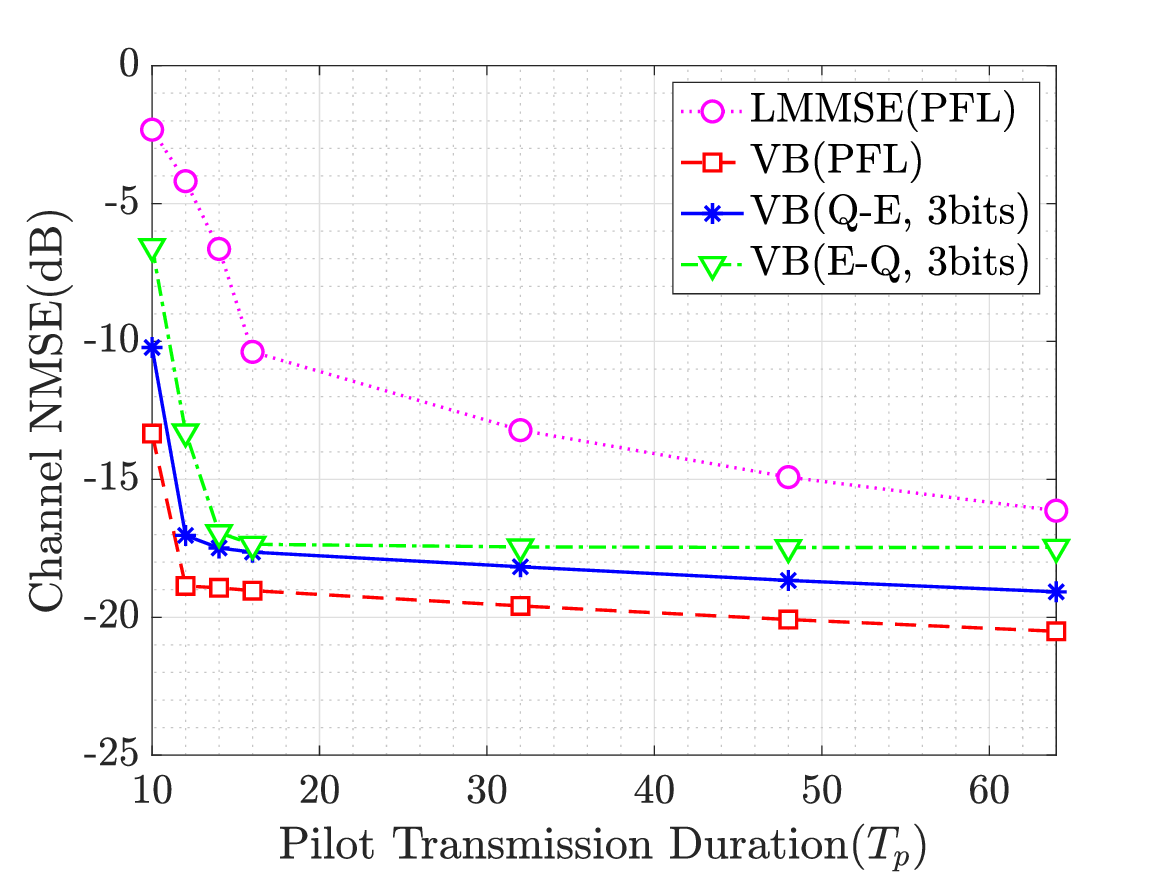}
  \vspace{-3mm}
  \caption{The channel NMSE performance comparison between LMMSE(PFL), VB(PFL), VB(Q-E, 3bits), and VB(E-Q, 3bits), when $L = 8, M = 4, K = 16, T_d=128$, $\text{SNR}=10$ dB, and $T_p\in [10,64]$.}\label{Fig:NMSE_vs_Tp}
  \vspace{-5mm}
\end{figure}

\begin{figure}[t]
\vspace{-6mm}
\centering
  \includegraphics[trim = 0mm 0mm 0mm 0mm, clip, scale=7, width=0.83\linewidth, draft=false]{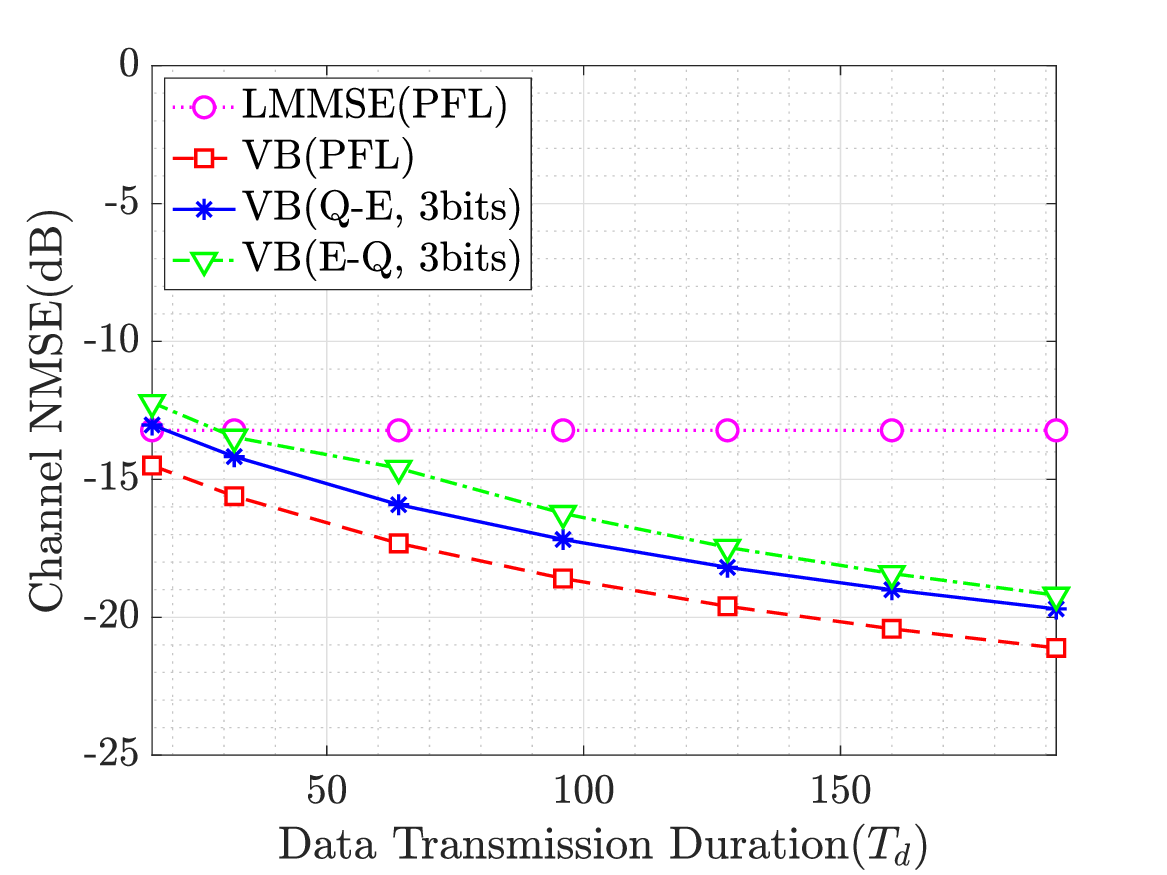}
  \vspace{-2mm}
  \caption{A comparison of channel NMSE among LMMSE(PFL), VB(PFL), VB(Q-E, 3bits), and VB(E-Q, 3bits) employing QPSK where $L = 8, M = 4, K = 16, T_p = 32$, $\text{SNR}=10$ dB, and $T_d$ ranging from $16$ to $192$.
  }\label{Fig:NMSE_vs_Td}
  \vspace{-5mm}
\end{figure}

Next, in Fig.~\ref{Fig:NMSE_vs_Tp} and Fig.~\ref{Fig:NMSE_vs_Td}, we study the channel NMSE performance across different values of $T_p$ and $T_d$, respectively, where the simulation settings are the same as in Fig.~\ref{Fig:SER_vs_Tp} and Fig.~\ref{Fig:SER_vs_Td}. In these figures, we compare LMMSE(PFL) with VB(PFL), VB(Q-E, 3bits), and VB(E-Q, 3bits). The results are similar to those in Fig. \ref{Fig:SER_vs_Tp} and Fig. \ref{Fig:SER_vs_Td}, respectively, where VB(PFL) outperforms the others. This observation shows that the CE and data detection procedures interactively affect each other, and improving one leads to improvements in the other.

\subsection{Computational Complexity}
In this subsection, we focus on the comparative analysis of computational complexity between LMMSE(PFL), VB(PFL), GAMP(PFL), VB-EM(PFL), VB(Q-E), GAMP(Q-E), VB-EM(Q-E), and VB(E-Q). In particular, the complexity of LMMSE(PFL) is $\mc{O}(MLK^2 + |\mc{S}|K)$ \cite{nguyen2022variational}. Moreover, the complexity of VB(PFL) is expressed as
$\mc{O}(I_{\mr{tr}}\big[M^3 L^3K + (T_d + 1)MLK + |\mc{S}|K\big])$, due to the matrix inversion in \eqref{Sigma-h} and matrix multiplications in \eqref{eq:q(gamma_p)} and \eqref{eq:q(gamma_d)}. The complexity of GAMP(PFL) and VB-EM(PFL) matches that of VB(PFL) \cite{wen2015bayes,nguyen2024variational}. Then, the complexity of VB(Q-E) is $\mc{O}(I_{\mr{tr}}\big[M^3 L^3K + (T_d + 1)MLK + |\mc{S}|K\big])$. Furthermore, the complexity of GAMP(Q-E) and VB-EM(Q-E) is equal to the complexity of VB(Q-E). Finally, the complexity of VB(E-Q) is given by
$\mc{O}(I_{\mr{tr}}\big[2M^3 LK + (T_d + 2)MK + |\mc{S}|K\big])$, due to the matrix inversions in \eqref{Sigma-h_E&Q} and \eqref{Sigma-h_CPU_E&Q}, and matrix multiplications in \eqref{eq:q(gamma_p)_E&Q} and within the CPU process. As a result, LMMSE(PFL) has the lowest complexity, followed by VB(E-Q), which has the second-lowest complexity, while the other methods based on PFL and Q-E exhibit the highest complexity.

%From these complexity expressions, we observe that the LMMSE(PFL) method has the lowest complexity, while the VB(PFL) and Q-E methods have the highest complexity. This highlights the trade-off between higher computational complexity and enhanced performance in the VB(PFL) and Q-E methods.

\subsection{Fronthaul Signaling Overhead}
\label{subsect:fronthaul_signaling}
In this part, we analyze the fronthaul signaling overhead requirements of the methods based on PFL, Q-E, and E-Q, as discussed in the previous subsection. In real-world scenarios, the unquantized signals at the APs must first be quantized (with high resolution) before being forwarded to the CPU. For this purpose, we assume a $10$-bit resolution to represent the unquantized signals. In this case, the PFL-based methods require $10ML(T_p + T_d)$ bits for fronthaul signaling overhead. In contrast, the VB(Q-E, 3bits), GAMP(Q-E, 3bits), and VB-EM(Q-E, 3bits) methods use 3-bit quantizers at the APs, where each AP quantizes its received signals and forwards them to the CPU, requiring $3ML(T_p + T_d)$ bits. On the other hand, VB(E-Q, 3bits) involves local CE at each AP, followed by quantization and forwarding of the quantized channels to the CPU, requiring $3ML(K+T_d)$ bits. This is lower than the methods based on (Q-E, 3bits) as $K\leq T_p$. %Therefore, if there is no issue with the bandwidth of fronthaul links, VB(PFL) is the optimal choice. Otherwise, the VB(Q-E, 3bits) method is more advantageous, as it provides better SER and channel NMSE performance compared to both the LMMSE(PFL) and VB(Q-E, 3bits) methods while requiring less fronthaul signaling overhead.

Based on the previous two subsections, we observe that, from a computational complexity perspective, among the VB-based methods, VB(PFL) and VB(Q-E) have the highest complexity. On the other hand, in terms of fronthaul signaling overhead, VB(PFL) requires the highest fronthaul load, and VB(Q-E, 3bits) imposes a higher load than VB(E-Q, 3bits). Therefore, there is a trade-off among the different VB-based methods. Specifically, if fronthaul bandwidth is not a concern, VB(PFL) is the optimal choice. If computational complexity is not a constraint but fronthaul bandwidth is limited, the VB(Q-E, 3bits) method is more favorable. Finally, if both computational complexity and fronthaul bandwidth are limited, the VB(E-Q, 3bits) method is the most suitable option.
        %%%%%%%%%%%%%%%%%%%%%%%%%%%%%%%%%%%%%%%%%%%%%%%%%%%	
        
 	\section{Conclusion}
        \label{Sec:conclusion}
        In this work, we studied the JED problem in CF-mMIMO networks. We proposed a VB-based approach to address the limitations of fronthaul bandwidth by using low-bit quantizers at the APs. Specifically, we considered two scenarios: Q-E and E-Q. In the Q-E scenario, each AP sends a quantized version of its received signals to the CPU, while in the E-Q scenario, each AP initially performs local CE during the pilot transmission phase and then shares a quantized version of the estimated channels, along with a quantized version of the received signals during the data transmission phase, with the CPU. %We utilized the VB framework to establish a nonlinear relationship between the quantized signals and their unquantized counterparts. We also treated the residual inter-user interference as a random variable and let the VB framework estimate its precision.
We evaluated our VB-based approach under PFL, Q-E, and E-Q scenarios and compared its performance with LMMSE(PFL), GAMP(PFL), GAMP(Q-E), VB-EM(PFL), and VB-EM(Q-E). Notably, VB(Q-E) and VB(E-Q) outperform LMMSE(PFL) due to the nonlinear VB framework. In addition, our VB-based methods outperform the nonlinear benchmarks thanks to their efficient convergence to a local optimum and the superior performance of the pure VB framework compared to approaches that rely on the EM technique. Future research can proceed in several promising directions. One direction is to explore the impact of near-field communication effects on the JED process in CF-mMIMO systems. Moreover, in any Bayesian-based statistical CE approach, including VB methods, knowledge of the channel covariance matrix is essential for effective estimation. This information can, in principle, be learned within the VB framework and will be considered an important direction for future work.
        %%%%%%%%%%%%%%%%%%%%%%%%%%%%%%%%%%%%%%%%%%%%%%%%%%%	
        %\section{Acknowledgment}
        % \label{Sec:Acknowledgment}
        % \subfile{Sections/Acknowledgment}
        %%%%%%%%%%%%%%%%%%%%%%%%%%%%%%%%%%%%%%%%%%%%%%%%%%%

%         \section*{Acknowledgment}
%        The authors would like to thank the anonymous reviewers for their insightful and constructive comments, which helped to improve the technical content and clarity of the paper. 
 
\appendix 
In this section, we provide details on how to compute the $\ms{U}$ and ${\sf V}$ functions for a truncated Gaussian distribution. To simplify the derivation, we define $x_1 = \sqrt{2\gamma}(a-\mu)$ and $x_2 = \sqrt{2\gamma}(b-\mu)$. Consider a random variable that follows a complex Gaussian distribution $\mathcal{CN}(\mu, \gamma^{-1})$, with its real and imaginary components truncated to the interval $(a,b)$. The expressions for the mean, $\ms{U}(\mu, \gamma, a, b)$, and variance, ${\sf V}(\mu, \gamma, a, b)$, can be obtained from \cite{nguyen2024variational}:
\begin{align}
    &\ms{U} (\mu,\gamma,a,b) = \mu - \frac{1}{\sqrt{2\langle \gamma \rangle}}\frac{u(x_2) - u(x_1)}{U(x_2) - U(x_1)},\label{eq:F_appndx}\\
    \label{eq:G_appndx}
    &\ms{V} (\mu,\gamma,a,b)  \\ 
    & \quad = \frac{1}{2\lr{\gamma }}\bigg[1 \!- \!\frac{x_2 u(x_2) \!- \!x_1 u(x_1)}{U(x_2) \!- U(x_1)} \!-\! \Big(\frac{u(x_2) - u(x_1)}{U(x_2) - U(x_1)} \Big)^2 \bigg]. \nonumber
\end{align}

	\bibliographystyle{IEEEtran}
        \bibliography{refs}

\end{document}